\documentclass[aps,prd,superscriptaddress,nofootinbib,floatfix,preprint]{revtex4-1}

\pdfoutput=0
\usepackage{graphicx}
\usepackage{mediabb}
\usepackage{amsmath}
\usepackage{bm}
\usepackage[small,bf]{caption}
\usepackage[subrefformat=parens]{subcaption}
\usepackage{comment}

\newcommand{\beq}{\begin{equation}}
\newcommand{\eeq}{\end{equation}}
\newcommand{\beqa}{\begin{eqnarray}}
\newcommand{\eeqa}{\end{eqnarray}}
\newcommand{\bpr}{\begin{problem}}
\newcommand{\epr}{\end{problem}}
\newcommand{\bcent}{\begin{center}}
\newcommand{\ecent}{\end{center}}
\newcommand{\bfig}{\begin{figure}}
\newcommand{\efig}{\end{figure}}
\newcommand{\bpc}{\begin{picture}}
\newcommand{\epc}{\end{picture}}
\newcommand{\barr}{\begin{array}}
\newcommand{\earr}{\end{array}}
\newcommand{\bitm}{\begin{itemize}}
\newcommand{\eitm}{\end{itemize}}
\newcommand{\bright}{\begin{flushright}}
\newcommand{\eright}{\end{flushright}}
\newcommand{\bminip}{\begin{minipage}}
\newcommand{\eminip}{\end{minipage}}
\newcommand{\btab}{\begin{tabular}}
\newcommand{\etab}{\end{tabular}}

\newcommand{\nnb}{\nonumber}

\newcommand{\hiroshima}{Graduate School of Advanced Science and Engineering, Hiroshima University, Kagamiyama, Higashi-Hiroshima 739-8526, Japan}

\newcommand{\icr}{Institute for Chemical Research, Kyoto University Uji, Kyoto 611-0011, Japan}

\newcommand{\kyoto}{Graduate School of Science, Kyoto University, Sakyouku, Kyoto 606-8502, Japan}
\newcommand{\ELINP}{Extreme Light Infrastructure-Nuclear Physics (ELI-NP)/Horia Hulubei National Institute for R\&D in Physics and Nuclear Engineering (IFIN-HH), 30 Reactorului St., P.O. Box MG-6, Bucharest-Magurele, Judetul Ilfov, RO-077125, Romania}
\newcommand{\NILPR}{National Institute for Laser, Plasma and Radiation Physics, 409 Atomistilor  PO Box MG-36, 077125, Magurele, Jud. Ilfov, Romania}

\newcommand{\om}{\omega}
\newcommand{\vth}{\vartheta}

\newcommand{\mcal}[1]{\mathcal{#1}}
\newcommand{\mrm}[1]{\mathrm{#1}}

\newcommand{\al}{\alpha}
\newcommand{\be}{\beta}

\newcommand{\ve}{\varepsilon}
\newcommand{\la}{\lambda}

\begin{document}
\title{Search for sub-eV axion-like resonance states via stimulated quasi-parallel laser collisions 
with the parameterization including fully asymmetric collisional geometry}

\author{Kensuke Homma\footnote{co-first author (corresponding author)}}\affiliation{\hiroshima}
\author{Yuri Kirita\footnote{co-first author}}\affiliation{\hiroshima}
\author{Masaki Hashida}\affiliation{\icr}\affiliation{\kyoto}
\author{Yusuke Hirahara}\affiliation{\hiroshima}
\author{Shunsuke Inoue}\affiliation{\icr}\affiliation{\kyoto}
\author{Fumiya Ishibashi}\affiliation{\hiroshima}
\author{Yoshihide Nakamiya}\affiliation{\icr}\affiliation{\ELINP}
\author{Liviu Neagu}\affiliation{\ELINP}\affiliation{\NILPR}
\author{Akihide Nobuhiro}\affiliation{\hiroshima}
\author{Takaya Ozaki}\affiliation{\hiroshima}
\author{Madalin-Mihai Rosu}\affiliation{\ELINP}
\author{Shuji Sakabe}\affiliation{\icr}\affiliation{\kyoto}
\author{Ovidiu Tesileanu}\affiliation{\ELINP}
\collaboration{SAPPHIRES collaboration}

\date{\today}

\begin{abstract}
We have searched for axion-like resonance states by colliding optical photons in
a focused laser field (creation beam) by adding another laser field (inducing beam) 
for stimulation of the resonance decays, where frequency-converted signal photons
can be created as a result of stimulated photon--photon scattering via exchanges of axion-like resonances.
A quasi-parallel collision system (QPS) in such a focused field allows access to the sub-eV mass range
of resonance particles.
In past searches in QPS, for simplicity, we interpreted the scattering rate based on 
an analytically calculable symmetric collision geometry 
in both incident angles and incident energies by partially implementing the
asymmetric nature to meet the actual experimental conditions.
In this paper, we present new search results based on a complete parameterization
including fully asymmetric collisional geometries. 
In particular, we combined a linearly polarized creation laser and a circularly polarized inducing laser
to match the new parameterization. 
A 0.10-mJ/31-fs Ti:sapphire laser pulse and a 0.20-mJ/9-ns Nd:YAG laser pulse were 
spatiotemporally synchronized by sharing a common optical axis and focused into the vacuum system. 
Under a condition in which atomic background processes were completely negligible,
no significant scattering signal was observed at the vacuum pressure of $2.6 \times 10^{-5}$~Pa,
thereby providing upper bounds on the coupling--mass relation 
by assuming exchanges of scalar and pseudoscalar fields at a 95\% confidence level in the sub-eV mass range.
%
%
%
\end{abstract}

\maketitle

\section{Introduction}
Spontaneous symmetry breaking is the key concept for understanding the fundamental laws of physics.
In particular, when a symmetry is global, the appearance of a massless Nambu--Goldstone boson (NGB)~\cite{NGB}
may be expected as a result of the broken symmetry. This viewpoint can be a robust guiding principle
for predicting new particle states based on various types of global symmetries 
in different theoretical contexts, including axion~\cite{axion}, dilaton~\cite{dilaton},
inflaton~\cite{miracle}, and string-inspired models~\cite{strings}.
However, NGBs are observed as pseudo-NGB states (pNGB) with finite masses due to complicated quantum corrections,
such as pions in the context of quantum chromodynamics (QCD). 
If pNGBs are coupled only very weakly to matter, they could be natural candidates to 
explain the dark components of the universe~\cite{InvAxion1,InvAxion2,InvAxion3}. 
When pNGB masses are relatively small
and the couplings to matter are feeble, they are referred to herein as axion-like particles (ALPs).
Although ALP masses are supposed to be small, how small is not known theoretically.
Therefore, experimental efforts to search for ALPs over wide low-mass and weak-coupling domains are valuable
for discovering some of the dark components of the universe.

The XENON1T experiment recently reported an excess of electron recoil events compared with
the defined background level ~\cite{XENON1T}.
Among three possible scenarios to explain this excess, the solar axion interpretation is 
the one that fits most nicely with the recoiled electron energy spectrum with 3.5~$\sigma$ significance.
The consistent axion mass range is $\mathcal{O}$(0.1--10)~eV in the photon--axion coupling range
of $\mathcal{O}$($10^{-12}$--$10^{-9}$)~GeV${}^{-1}$ depending on the electron--axion coupling
and the QCD axion models~\cite{DFSZ,KSVZ}.
However, there is strong tension between this interpretation and the existing constraints 
from stellar cooling~\cite{Stellar1,Stellar2,Stellar3,Stellar4,Stellar5}. If the aforementioned significance level further increases 
through improved observations in the near future and the solar axion scenario remains the one that is most valid, 
then this tension becomes a real issue. 
To resolve this issue, model-independent observations of the direct production 
of axions and their decay in laboratory experiments would be indispensable.

In this paper, we focus on the coupling between sub-eV ALPs and laser photons.
To describe the coupling of scalar ($\phi$) or pseudoscalar ($\sigma$) ALPs
to two photons, the following two effective Lagrangians are considered: 
\beq
-L_{\phi} = gM^{-1}\frac{1}{4}F_{\mu\nu}F^{\mu\nu}\phi , \hspace{10pt} -L_{\sigma} = gM^{-1}\frac{1}{4}F_{\mu\nu}\tilde{F}^{\mu\nu}\sigma,
\label{eq_phisigma} 
\eeq
where 
$g$ is a dimensionless constant for a given energy scale $M$ at which 
a relevant global symmetry is broken,
and
$F^{\mu\nu}=\partial^{\mu}A^{\nu}-\partial^{\nu}A^{\mu}$ is the electromagnetic 
field strength tensor
and its dual $\tilde{F}^{\mu\nu} \equiv \ve^{\mu\nu\alpha\beta}F_{\alpha\beta}$ with
the Levi-Civita symbol $\ve^{ijkl}$.

\begin{figure}[h]
\begin{center}
\includegraphics[scale=0.48]{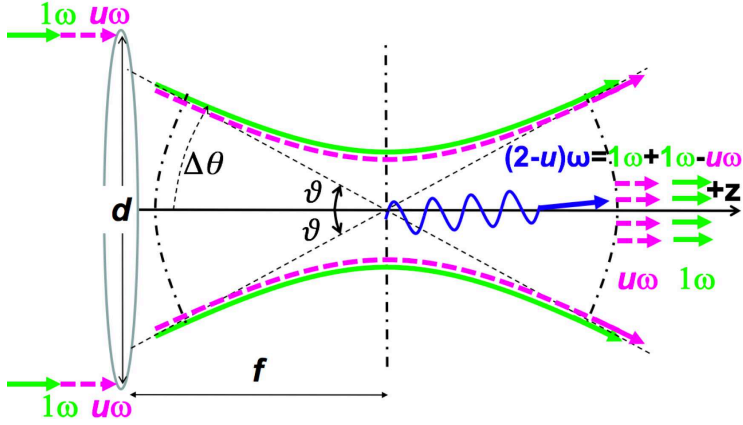}
\end{center}
\caption{
Concept of stimulated resonant photon--photon scattering in a quasi-parallel collision system (QPS)
by focusing two-color laser fields in a vacuum.
This figure is quoted from~\cite{PTEP-EXP00} with a slight modification.
A coherent field with energy $\omega$ (solid green line) is combined with a different-color coherent field
with energy $u\omega~(0<u<1)$ (dashed red line). The combined fields are focused by a lens element
in a vacuum. Emission of signal photons with energy $(2-u)\omega$ (blue wavy line) is 
stimulated as a result of energy--momentum conservation in the scattering process
$\omega + \omega \rightarrow \phi/\sigma \rightarrow (2-u)\omega + u\omega$ via a resonance
state $\phi/\sigma$. 
Given the focusing parameters of beam diameter $d$ and focal length $f$, 
the incident angle $\vartheta$ is expected to vary over $0<\vartheta\leq \Delta\theta$.
This unavoidable ambiguity of the wave vectors of the incident light waves provides a wide window
for accessing different center-of-mass system energies at a given point in time.}
\label{Fig1}
\end{figure}

Focusing on sub-eV ALPs, we have proposed to utilize quasi-parallel collision system (QPS)
between two photon pairs with equal energy $\omega$ by combining and focusing two-color lasers 
along a common optical axis~\cite{DEptp} as illustrated in Fig.~\ref{Fig1}.
The corresponding center-of-mass system (CMS) energy in the QPS is expressed as
\beq
E_{CMS}=2\omega\sin{\vartheta},
\eeq
where $2\vartheta$ is the relative angle between a pair of incident photons.
By controlling the beam diameter ($d$) and focal length ($f$) experimentally, the QPS can be
sensitive to ALP resonance states with the mass range of $0 < m < 2\omega\sin\Delta\vartheta$, 
where $m$ is the ALP mass and $\Delta\vartheta$ can be adjusted by the focusing geometry determined 
with $\Delta\vartheta \sim (d/2)/f$.

The first key feature of the proposed method~\cite{DEptp} is the resonant ALP production via the $s$-channel
exchange within the $E_{CMS}$ uncertainty, which drastically enhances the production rate~\cite{DEptp}.
The second key feature is stimulated decays of produced ALPs
to fixed final states via energy--momentum conservation between four photons in the initial and
final states. This stimulated resonant scattering rate eventually becomes proportional to
the square of the number of photons in the creation laser beam and to the number of photons
in the inducing laser beam. This cubic dependence on the number of photons in the beams 
offers opportunities to search for ALPs with extremely weak coupling 
when the beam intensity is high enough~\cite{JHEP}.

\begin{figure}[h]
\begin{tabular}{cc}
 \begin{minipage}[t]{0.49\linewidth}
  \centering
  \subcaption{}\label{Fig2a}
  \includegraphics[keepaspectratio, scale=0.38]
  {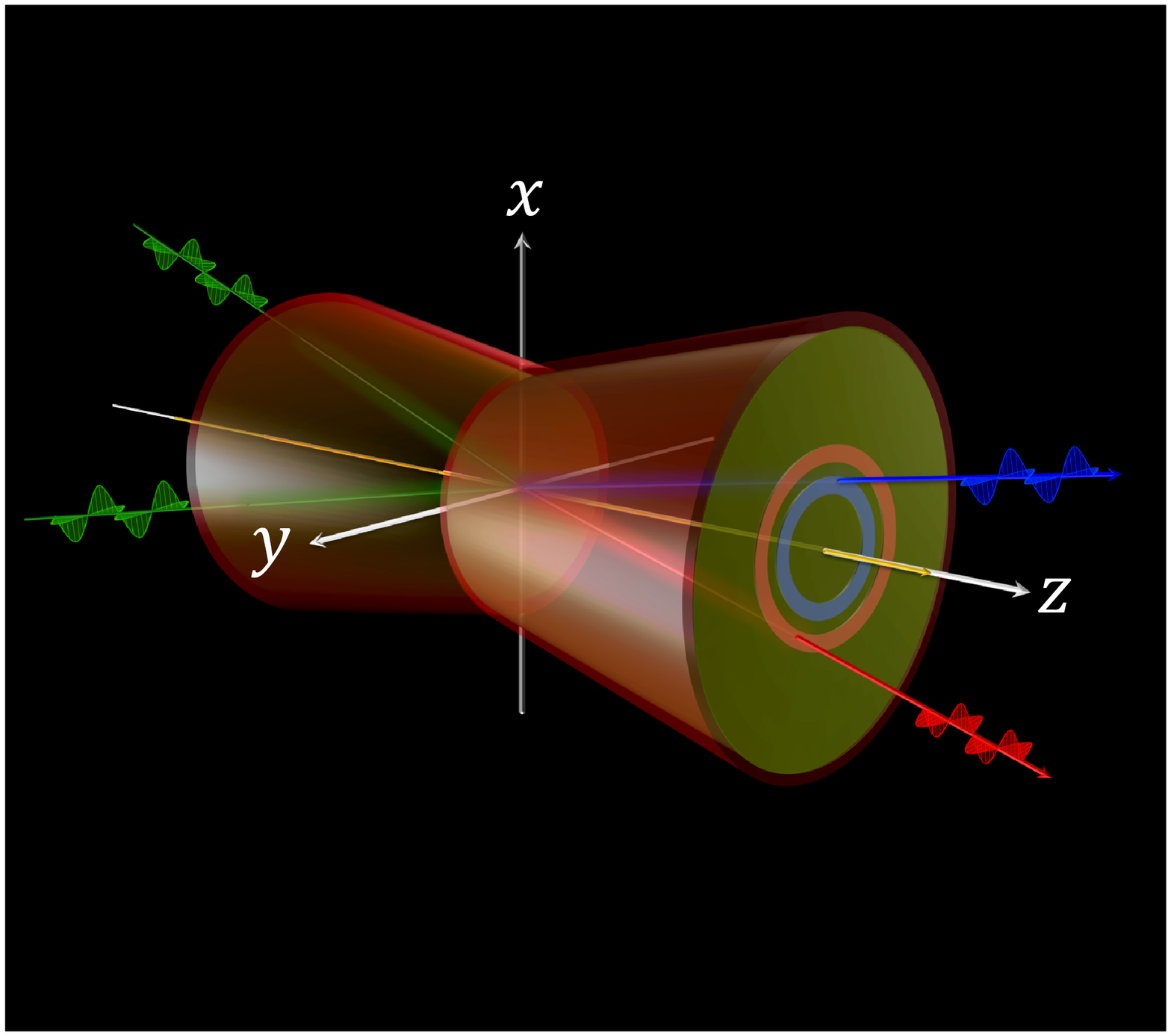}
 \end{minipage}
&
 \begin{minipage}[t]{0.49\linewidth}
  \centering
  \subcaption{}\label{Fig2b}
  \includegraphics[keepaspectratio, scale=0.38]
  {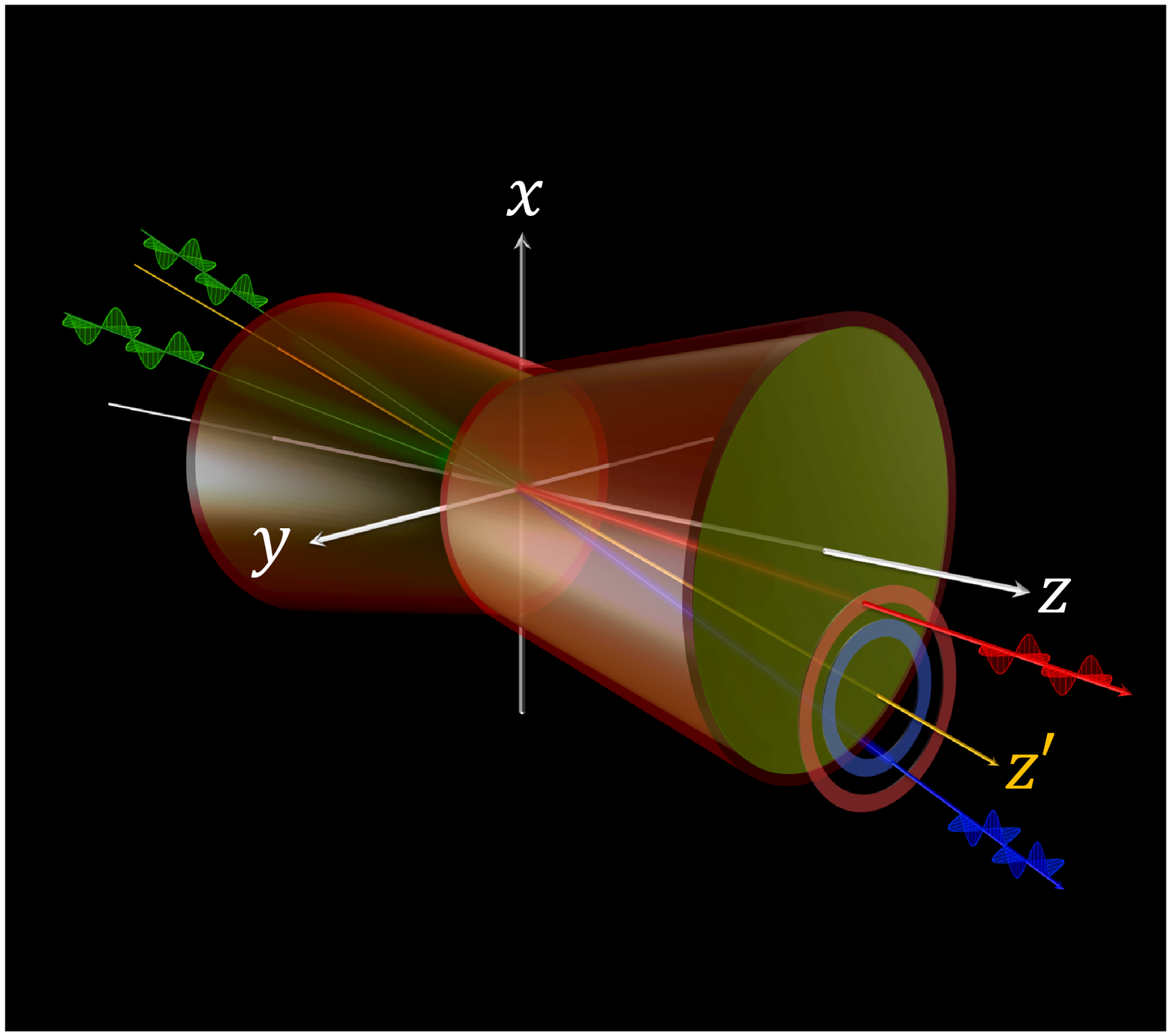}
 \end{minipage}
\end{tabular}
\caption{
Classification of collision geometries in QPS. The two figures are extracted from Fig.1 of~\cite{JHEP}.
(a) Symmetric incidence and coaxial scattering, where the incident angles of the two wave vectors
and their energies are symmetric, and the transverse momentum of the photon pair, $p_T$,
always vanishes with respect to the common optical axis $z$.
(b) Asymmetric incidence and non-coaxial scattering, where the incident angles of the two wave
vectors and their energies are asymmetric, resulting in a finite value of
$p_T$ with respect to the common optical axis $z$. The zero-$p_T$ axis ($z^{'}$-axis) 
is always configurable for arbitrary pairs of incident wave vectors.
}
\label{Fig3}
\end{figure}

In past searches~\cite{PTEP-EXP00,PTEP-EXP01,PTEP-EXP02}, 
we provided constraints on the coupling--mass relation based on a symmetric QPS interpretation
in which the incident angles and energies of the two initial-state photons are symmetric (Fig.~\ref{Fig1}).
Based on this parameterization, we provided conservative constraints, and 
we respected the simple analytic treatment with symmetric collisions because the initial search was made
with narrow-bandwidth lasers.
However, in a short-pulse laser that is close to the Fourier-transform limit, where the relation between
laser frequency and time duration is governed by the wavelike nature of the system (i.e., the uncertainty principle),
we must accept an energy spread in principle, and so the approximation of symmetric incident energies is not realistic.
In addition, at the diffraction limit where the beam diameters reach their minimum values, 
the incident angles must also fluctuate greatly because of momentum--position uncertainties.
Therefore, we must accept a situation in which the incident photon energies and angles are both asymmetric.
Recently, we formulated the interaction rate based on the fully asymmetric collision system~\cite{JHEP}, where the non-coaxial geometry of the two-photon collisions and stimulated decays are 
explicitly included with respect to a given coaxial geometry of focused beams (Fig.~\ref{Fig2b}).

In general, linear polarization states in laser beams are supposed to be fixed precisely.
However, linear polarization states around a focal point are not simple plane-wave states
because the directions of the wave vectors are not identical in three dimensions, as illustrated in Fig.~\ref{Fig3},
even if the wave vectors are all aligned to a unique direction before focusing.
If we require the transverse wave states of individual light waves, then
the polarization directions of individual light waves must
fluctuate depending on the individual wave vector directions.
Therefore, these fluctuations must be included in both the creation and inducing lasers around the focal points.
In order for the stimulation due to the associated inducing field to be effective, one of the final-state
light waves must coincide with the momentum and also with the linear polarization state of the inducing
laser waves. Therefore, fluctuations of linear polarization directions make the evaluation of 
the stimulation effect in non-coaxial scattering events very complicated because coaxial symmetry 
of the inducing laser beam is no longer applicable.
However, if a collection of light waves is circularly polarized,
we can avoid this complication because even if the directions of the
waves in three dimensions are changed, the individual light waves retain the circular polarization.
Therefore, in~\cite{JHEP} we evaluated the stimulation effect based on circularly polarized laser beams. 
However, there is an experimental constraint in
that high-intensity laser pulses are provided as linearly polarized states, and 
changing their polarization from linear to circular adds a technical difficulty. 
In this paper, we report new results for sub-eV ALP searches by combining
linearly and circularly polarized states for the creation and inducing lasers, respectively,
based on the parameterization including the fully asymmetric collision geometry in QPS.
This is in contrast to previous works in which we reported searches based on the symmetric QPS approximation
with linear polarization states for both the creation and inducing lasers~\cite{PTEP-EXP00,PTEP-EXP01,PTEP-EXP02}.

\section{Formulae for obtaining $m$--$g/M$ relation numerically in general QPS geometry}
Stimulated resonant photon--photon scattering in the most general
collisional geometry including asymmetric incidence and non-coaxial scattering was formulated in~\cite{JHEP}, and 
the full details can be found in the appendix of~\cite{JHEP}.
In the following subsections, we briefly explain how to relate the physical parameters of 
mass $m$ and coupling $g/M$ with the observed number of stimulated photon signals by reviewing 
only the relevant formulae to discuss the interaction rate dedicated for this search. 
We specifically replace the vertex factors in the scattering amplitude
because we must change from the circular polarization formulated in~\cite{JHEP}
to the linear one for both scalar and pseudoscalar field exchanges in this search.

We address a search for signal photons $p_3$ for the following degenerate case
in the generic QPS: 
\beq\label{eq_scattering}
<p_c(p_1)> + <p_c(p_2)> \rightarrow p_3 + <p_i(p_4)>,
\eeq
where $<\quad>$ indicates that $p_1$ and $p_2$ are chosen stochastically 
from a single focused coherent beam whose central four-momentum is $p_c$ 
for the creation of ALPs via $s$-channel photon--photon scattering, 
while the focused coherent beam with the central four-momentum $p_i$ is co-moving
to induce emission of signal photons $p_3$ when a fraction of the $p_i$ beam coincides with $p_4$.

In symmetric incidence and coaxial scattering as illustrated in Fig.~\ref{Fig2a},
transverse momenta of stochastically selected wave pairs, $p_T$, are constraint to 
be zero with respect to the common optical axis $z$.
Given that the azimuthal angles of the final-state wave vectors are axially symmetric
around the $z$-axis, the evaluation of the inducible momentum or angular range can be greatly simplified 
owing to the axial symmetric nature of the focused laser beams.
In contrast, asymmetric incidence and non-coaxial scattering in Fig.~\ref{Fig2b} introduces
a finite transverse momentum.
In this case, a new zero-$p_T$ axis, referred to as the $z^{'}$-axis, can be found 
for the pair of incident wave vectors. Based on the $z^{'}$-axis,
the axial symmetric nature of the azimuthal angles of the final-state wave vectors can be restored.
However, the inducing coherent field is still physically fixed to the common optical axis $z$.
This situation complicates the evaluation of the inducible momentum range 
depending on an arbitrarily formed $z^{'}$-axis. To solve this complication,
a numerical integration is required to express the number of signal photons
in the scattering process (\ref{eq_scattering}) per pulse collision, ${\cal Y}_{c+i}$, 
as soon reviewed in the following subsections.
With ${\cal Y}_{c+i}$ and a set of laser beam parameters $P$,
the number of stimulated signal photons, $N_{obs}$, as a function of
mass $m$ and coupling $g/M$ is eventually expressed as
\beq\label{Nobs}
N_{obs} = {\cal Y}_{c+i}(m, g/M ; P) t_{a} r \epsilon ,
\eeq
where $t_a$ is the data acquisition time, $r$ is the repetition rate of the pulsed beams, and
$\epsilon$ is the efficiency of detecting $p_3$.
For a set of $m$ values and an $N_{obs}$,
a set of coupling $g/M$ is evaluated by solving this equation numerically.

\subsection{Induced signal yield ${\cal Y}_{c+i}$}
With the average numbers of photons $N_c$ and $N_i$ for the creation and inducing coherent beams, respectively,
and units given in [\quad] with units of length $L$ and second $s$,
the induced signal yield ${\mathcal Y}_{c+i}$ per pulse collision is evaluated as
\beqa\label{eq_Yci}
{\mathcal Y}_{c+i}[1] = (N_c/2) (N_c/2) N_i \times \mbox{\hspace{4cm}} \\ \nnb
\left(
\int_{-Z_R/c}^0 dt \int_{-\infty}^{+\infty} d\bm{r} \rho_c(\bm{r},t) \rho_c(\bm{r},t) \rho_i(\bm{r},t) V_i
\right)
\times \mbox{\hspace{0.8cm}} \\ \nnb
\left(
\int dQ_I W(Q_I)
\frac{c}{2\om_1 2\om_2} |{\mathcal M}_s(Q^{'})|^2 dL^{'I}_{ips} \mbox{\hspace{0.1cm}}
\right)
\\ \nnb
\equiv  \frac{1}{4} N^2_c N_i {\mathcal D}_I\left[s/L^3\right] \overline{\Sigma}_I\left[L^3/s\right].
\mbox{\hspace{2.3cm}}
\eeqa

${\mathcal D}_I$ in Eq.(\ref{eq_Yci}) is a spatiotemporal overlapping factor in laboratory coordinates
(see $x,y,z$ in Fig.~\ref{Fig3}) of the focused creation beam (subscript $c$)
with the co-moving focused inducing beam (subscript $i$) limited in the Rayleigh length $Z_R$ 
only around the focal spot for a conservative evaluation.
The following photon number densities $\rho_{k=c,i}$ based on the Gaussian beam parameterization
are integrated over spacetime $(t, \bm{r})$: 
\beqa
\rho_{k}(t,\bm{r})
 = \left( \frac{2}{\pi} \right)^{\frac{3}{2}}\frac{1}{w_{k}^{2}(ct)c\tau_{k}}
\times \mbox{\hspace{3.4cm}} \\ \nnb
\exp\left( -2\frac{x^{2}+y^{2}}{w_{k}^{2}(ct)} \right)
\exp\left( -2\left( \frac{z-ct}{c\tau_{k}} \right)^{2} \right),
\eeqa
where $w_k$ are the beam radii as a function of time $t$ whose origin is set at the moment when all the pulses
reach the focal point, and $\tau_k$ are the time durations of the pulsed laser beams
with the speed of light $c$ and the volume of the inducing beam $V_i$ defined as
\beq
V_i = (\pi/2)^{3/2} w^2_{i0} c\tau_i ,
\eeq
where $w_{i0}$ is the beam waist (minimum radius) of the inducing beam.
The actually used overlapping factor configured for the case of different beam diameters between
creation and inducing lasers is summarized in the appendix of this paper.
 
$\overline{\Sigma}_I$ in Eq.(\ref{eq_Yci}) is an integrated inducible volume-wise interaction rate that
integrates the square of the scattering amplitude $|{\mathcal M}_s(Q^{'})|^2$ 
over an inducible variable set 
comprising energies $\om_i$, polar angles $\Theta_i$, and azimuthal angles 
$\Phi_i$ in laboratory coordinates for $i=1,2,4$:
$Q_I \equiv \{Q, \om_4, \Theta_4, \Phi_4\}$ 
with $Q \equiv \{\om_1, \Theta_1, \Phi_1, \om_2, \Theta_2, \Phi_2\}$
by weighting with multiple Gaussian distributions:
\beq
W(Q_I) \equiv \Pi_i G_E(\om_i)G_p(\Theta_i,\Phi_i)
\eeq
with Gaussian distributions on energy $G_E$ and momentum $G_p$, and 
also over an inducible Lorentz-invariant phase space in zero-$p_T$ coordinates:
\beq\label{eqLIPS}
dL^{'I}_{ips} = (2\pi)^4 \delta(p^{'}_3+p^{'}_4-p^{'}_1-p^{'}_2)
\frac{d^3p^{'}_3}{2\omega_3(2\pi)^3}\frac{d^3p^{'}_4}{2\omega_4(2\pi)^3}
\eeq
with two incident energies $\om_1$ and $\om_2$,
where the primed variables are converted from $Q$ in laboratory coordinates to $Q^{'}$ in zero-$p_T$ coordinates
via coordinate rotation.
The inducing weight $W(Q_I)$ takes care of the energy and momentum fractions of $p_4$ 
satisfying energy--momentum conservation with respect to the energy and momentum distributions
of the given inducing laser beam in laboratory coordinates.
The essential element of $\overline{\Sigma}_I$ is the Lorentz-invariant scattering amplitude defined
in zero-$p_T$ coordinates,
${\cal M}_S(Q^{'})$, for the given polarization states $S=abcd$ in a two-body interaction:
$p^{'}_1\{a\} + p^{'}_2\{b\} \rightarrow p^{'}_3\{c\} + p^{'}_4\{d\}$.
Unless confusion is expected, for simplicity, we omit the prime symbol associated 
with the momentum vectors in the following explanations.

\subsection{Vertex factors in scattering amplitude ${\cal M}_S$}
\begin{figure}[!h]
\begin{center}
\includegraphics[scale=0.7]{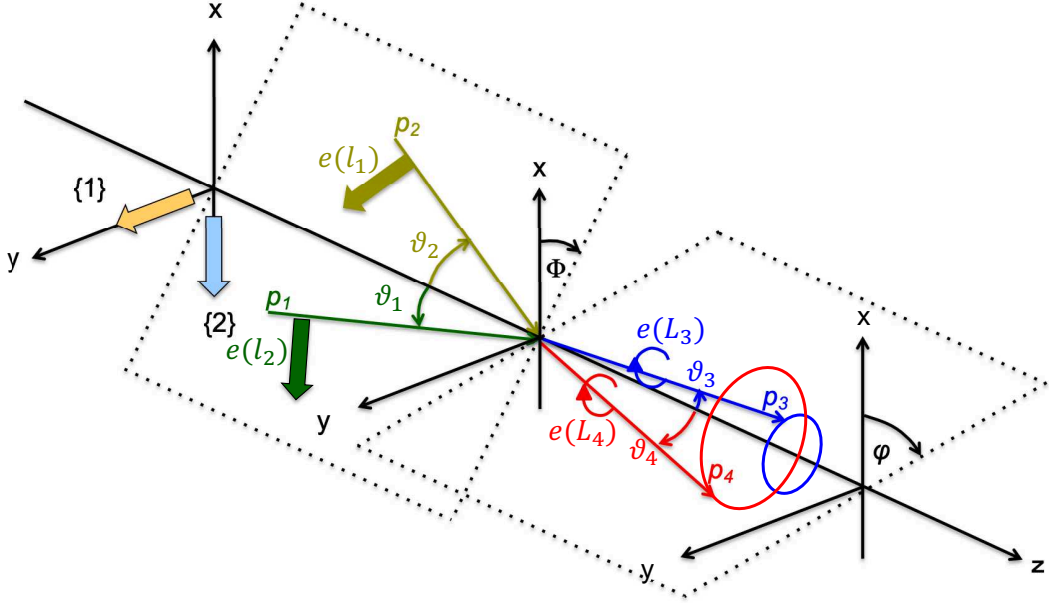}
\end{center}
\caption{
Definitions of four-momentum vectors $p_i$ and four-polarization vectors $e(\lambda_{p_i})$
with polarization states $\lambda_{p_i}$ for the initial-state ($i=1,2$) and final-state ($i=3,4$) light waves.
$l_i$ indicates mixture of experimentally defined orthogonal linear polarization states $\{1\}$ and $\{2\}$ 
before focusing in laboratory coordinates, and $L_i$ indicates left-handed circular polarization states.
These polarization vectors are mapped on the rotating reaction planes 
in the case of asymmetric incident angles and energies in QPS, 
where the transverse momentum $p_T$ of the $p_1+p_2$ pair is constraint to be zero 
as a specially simple case of general asymmetric collisions.
}
\label{Fig3}
\end{figure}

The polarization information is normally useful for distinguishing whether ALPs are scalar or
pseudoscalar fields.
When a two-body photon--photon scattering process
$p_1 + p_2 \rightarrow p_3 + p_4$ in four-momentum space
occurs on an identical reaction plane, namely, when the coplanar condition
($\Phi=\varphi=0$ in Fig.~\ref{Fig3}) is satisfied,
the difference between scalar and pseudoscalar cases becomes distinct.
Given an orthogonal set of linear polarization states $\{1\}$ and $\{2\}$,
the non-zero scattering amplitudes are limited to the following cases:
\beqa\label{eq_sc}
p_1\{1\} + p_2\{1\} \rightarrow p_3\{2\} + p_4\{2\} \nnb ,\\
p_1\{1\} + p_2\{1\} \rightarrow p_3\{1\} + p_4\{1\}
\eeqa
for scalar field exchange and
\beqa \label{eq_ps}
p_1\{1\} + p_2\{2\} \rightarrow p_3\{1\} + p_4\{2\} \nnb ,\\
p_1\{1\} + p_2\{2\} \rightarrow p_3\{2\} + p_4\{1\}
\eeqa
for pseudoscalar field exchange, 
where swapping $\{1\}$ and $\{2\}$ gives the same scattering amplitudes.

However, as illustrated in Fig.~\ref{Fig3}, 
the coplanar condition is, in most cases, not satisfied in QPS in contrast to CMS
because the $\vec{p}_1-\vec{p}_2$ plane and the $\vec{p}_3-\vec{p}_4$ plane
may differ from the $x$--$z$ plane defined with the laboratory coordinates.
Therefore, we must introduce combinations of linear polarization states, $l_1$ and $l_2$, based on the theoretically 
introduced planes with respect to the experimental linear polarization states \{1\} and \{2\},
which are mapped to the $y$ and $x$ axes, respectively.
In the following search, we assign the P-polarized state of the creation laser 
to the \{2\}-state and combine it with the circularly polarized inducing laser. 
Note here that due to the rotating nature of the incident reaction plane
in the focused geometry, even if the experimentally prepared linear polarization state is 
limited to \{2\}, polarization states defined on individual $\vec{p}_1-\vec{p}_2$ planes 
can contain both \{1\} and \{2\} components with different projection weights,
resulting in sensitivities to both scalar and pseudoscalar fields.
This situation is implemented quantitatively in the vertex factors as follows.

Based on expansion of the electromagnetic field strength tensor
$F^{\mu\nu}$ and its dual $\tilde{F}^{\mu\nu}$,
momentum-polarization tensors corresponding to the expanded coefficients
are defined (see Eqs.~(A.5) and (A.6) in \cite{JHEP} for details).
With a polarization four-vector $e_i(\la_p)$ with an arbitrary polarization state $\la_p$ 
associated with a four-momentum $p$, and the symbol * indicating the complex conjugate,
the momentum-polarization tensors are defined as
\beqa\label{eq_Tensor}
P^{\mu\nu}(\la_p) &\equiv& p^{\mu}e^{\nu}(\la_p)-e^{\mu}(\la_p)p^{\nu},
\\ \nnb
\hat{P}^{\mu\nu}(\la_p) &\equiv&  e^{*\mu}(\la_p)p^{\nu}-p^\mu e^{*\nu}(\la_p)
\eeqa
for the tensor $F^{\mu\nu}$
and
\beqa\label{eq_TensorT}
\tilde{P}^{\mu\nu}(\la_p) &\equiv& \ve^{\mu\nu\al\be} \left( p_{\al}e_{\be}(\la_p) - e_{\al}(\la_p)p_{\be} \right),
\\ \nnb
\hat{\tilde{P}}^{\mu\nu}(\la_p) &\equiv& \ve^{\mu\nu\al\be} \left( p_{\al}e^{*}_{\be}(\la_p) - e^{*}_{\al}(\la_p)p_{\be} \right)
\eeqa
for the dual tensor $\tilde{F}^{\mu\nu}$.

Given the vector and tensor definitions above, the Lorentz-invariant scattering amplitude ${\cal M}_S$ 
dedicated for scalar field exchange is expressed as 
(see Eq.~(A.33) in \cite{JHEP} for the detailed derivation)
\beqa
{\cal M}_S = \frac{1}{4} \left(\frac{g}{M}\right)^2
\frac{(P_{1}P_{2})(\hat{P}_{3}\hat{P}_{4})}{m^2-(p_1+p_2)^2},
\eeqa
where the factors $(P_i P_j)$ in the numerator correspond to the vertex factors 
reflecting polarization states in the initial and final states, respectively.
$(ST)$ is the abbreviation for a momentum-polarization tensor product such as
$(ST) \equiv S_{\mu\nu}T^{\mu\nu}$ for four-momenta $s$ and $t$, that is,
$(P_1 P_2)$ corresponds to a momentum-tensor product for four-momenta $p_1$ and $p_2$.
In the case of pseudoscalar exchange, we have only to replace the vertex factors
with $(P_1 \tilde{P_2})(\hat{P_3} \hat{\tilde{P_4}})$ using Eq.~(\ref{eq_TensorT}).

Hence, necessary momentum-polarization tensor products between four-momenta $s$ and $t$ 
with their polarization states $\la_s$ and $\la_t$ are summarized as
\beqa\label{eq_ST}
S_{\mu\nu}(\la_s) T^{\mu\nu}(\la_t) &=& 
2\{(s \cdot t)(e(\la_s) \cdot e(\la_t)) - (s \cdot e(\la_t)) (t \cdot e(\la_s))\},
\\ \nnb
\hat{S}_{\mu\nu}(\la_s) \hat{T}^{\mu\nu}(\la_t) &=& 
2\{(s \cdot t)(e^{*}(\la_s) \cdot e^{*}(\la_t)) - (s \cdot e^{*}(\la_t)) (t \cdot e^{*}(\la_s))\}
\eeqa
for scalar field exchange and
\beqa\label{eq_STT}
S_{\mu\nu}(\la_s) \tilde{T}^{\mu\nu}(\la_t) &=& 
4 \ve^{\mu\nu\al\be} s_{\mu} e_{\nu}(\la_s) t_{\al} e_{\be}(\la_t),
\\ \nnb
\hat{S}_{\mu\nu}(\la_s) \hat{\tilde{T}}^{\mu\nu}(\la_t) &=& 
4 \ve^{\mu\nu\al\be} s_{\mu} e^{*}_{\nu}(\la_s) t_{\al} e^{*}_{\be}(\la_t)
\eeqa
for pseudoscalar exchange.

The actually used vertex factors for scalar and pseudoscalar field exchanges 
dedicated for this search with the fixed left-handed circular polarization state, $L$, 
of the inducing laser are expressed as follows:
\beqa\label{helicity}
\mbox{scalar type}&:& ({P_1}(l_1) {P_2}(l_2)) (\hat{{P_3}}(L_3) \hat{{P_4}}(L_4));
\\ \nnb
\mbox{pseudoscalar type}&:& ({P_1}(l_1) \tilde{{P_2}}(l_2)) (\hat{{P_3}}(L_3) \hat{\tilde{{P_4}}}(L_4)), 
\eeqa
where $l_i$ with $i=1,2$ represents mixing of linear polarization states \{1\} and \{2\}
due to rotation of the $p_1-p_2$ reaction plane with respect to the linear polarization
direction of the creation laser beam, while rotation of the $p_3-p_4$ reaction plane
does not affect the circular polarization states of photons due to helicity conservation.
We address only the $L$-state for $p_3$~$(L_3)$ induced by the $L$-state for $p_4$~$(L_4)$ 
in the inducing field.  This is because vertex factors combining opposite
circular polarization states always vanish, counter-intuitively, in both scalar and pseudoscalar exchanges
based on Eqs.~(\ref{eq_ST}) and (\ref{eq_STT}).

\begin{figure}[!hbt]
\centering
\includegraphics[width=0.80\textwidth]{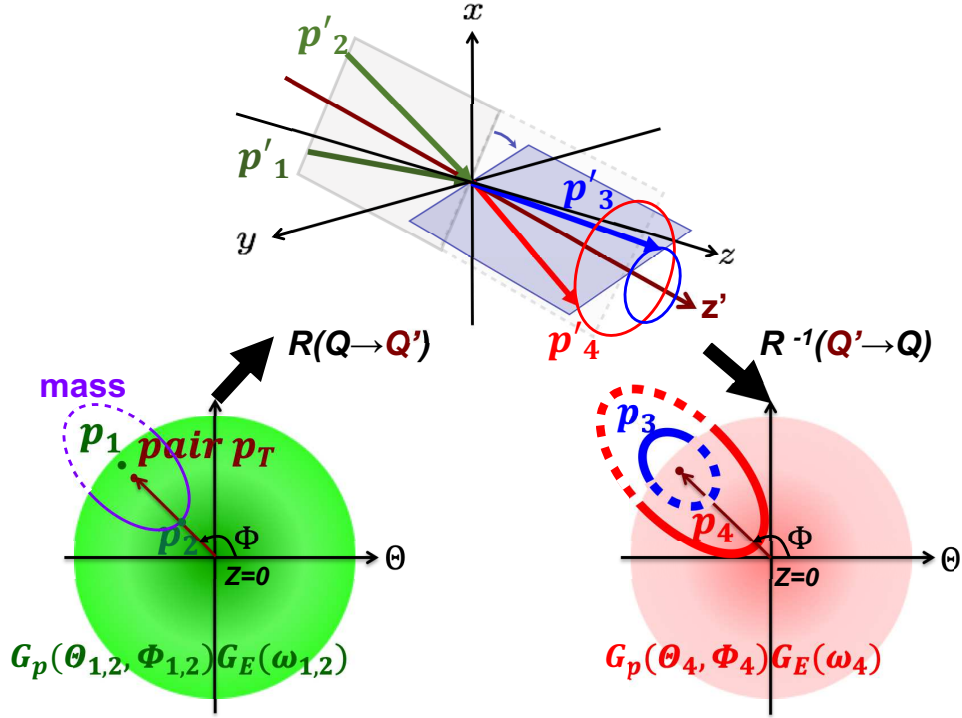}
\caption{Flow of numerical calculations. The details are explained in the main text.}
\label{Fig4}
\end{figure}

\subsection{Outline of numerical calculations}
Figure~\ref{Fig3} is a special example of an energy-incident-angle asymmetric collision that accidentally shares the common
optical axis of the two laser beams, namely, the incident $p_T$ is set to zero.
The actual procedures in the numerical calculation including fully asymmetric collisions
are explained below based on the illustration in Fig.~\ref{Fig4}. 
We first introduce momentum distributions $G_p$ as a function of polar angles $\Theta_i$ and 
azimuthal angles $\Phi_i$ mapped on the laboratory coordinates and 
energy distributions $G_E(\omega_i)$ for the creation (left, green)
and inducing (right, red) lasers for individual photons $i=1,2,4$ by denoting $G$ representing 
normalized Gaussian distributions.
Angular distributions in cylindrical coordinates $(\Theta, \Phi, z=0)$ (diameter corresponds to $\Theta$)
are used here representing $G_p$ by fixing momentum norms at energies chosen from $G_E$.
The concrete steps are then as follows.
Step~0: Select a finite-size segment of $p_1$ based on the $G_E(\omega_1) G_p(\Theta_1,\Phi_1)$ distributions.
Step~1: A $z^{'}$-axis of zero-$p_T$ coordinates is defined
by finding a paring $p_2$ that satisfies the resonance condition 
with respect to the selected $p_1$ and to a finite energy segment from $G_E(\om_2)$
for a given mass parameter $m$. The kinematically possible ellipsoidal orbit is drawn 
with the purple belt on the angular distribution.
Step~2: Convert the polarization vectors $e_i(\la_i) (i=1,2)$ from laboratory coordinates to zero-$p_T$ coordinates
through coordinate rotation $R(Q \rightarrow Q^{'})$.
Step~3: Calculate the invariant amplitude based on the vertex factors for scalar and pseudoscalar exchanges,
respectively, in the given zero-$p_T$ coordinates 
where the axial symmetric nature of the final-state $p^{'}_3$ and $p^{'}_4$ around $z^{'}$ is preserved.
Step~4: To evaluate the inducing effect
with respect to $G_E(\om_4)G_p(\Theta_4,\Phi_4)$ defined in laboratory coordinates,
a matching fraction of $p_4$ is calculated after rotating back to the laboratory coordinates
from the zero-$p_T$ coordinates, denoted by $R^{-1}(Q^{'} \rightarrow Q)$.
Based on the spread of $G_E(\om_4)$, the red ellipsoidal belt is determined via energy--momentum conservation.
Note here that due to the circular polarization state of the inducing beam, 
any $p_4$ experiencing scattering can satisfy the polarization matching to the $p_i$ beam.
Step~5: $p_3$ must balance with $p_4$ through energy--momentum conservation, so we can define 
a parametric signal energy spread via $\om_s \equiv \om_3 = \om_1+\om_2-\om_4$ as well as the polar and
azimuthal angle spreads by taking the $G_E(\om_4)G_p(\Theta_4,\Phi_4)$ distribution into account. 
The volume-wise interaction rate $\overline{\Sigma}_I$ is integrated over the inducible solid angle of $p_3$ calculated from all the energy and angular spreads.
Step~6: The signal yield ${\cal Y}_{c+i}$ is finally calculated based on Eq.~(\ref{eq_Yci}).

\begin{figure}[!h]
\begin{center}
\includegraphics[scale=0.8]{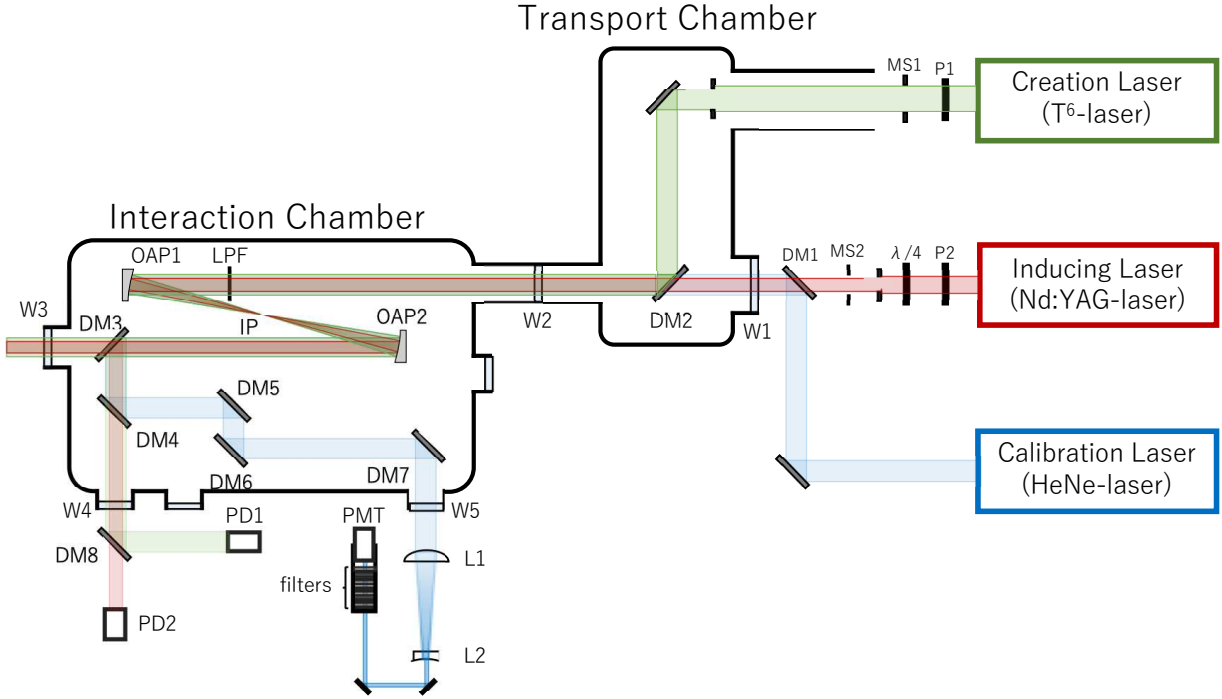}
\end{center}
\caption{Schematic of the searching system contained in the beam transport chamber
and interaction chamber designed to achieve $10^{-8}$~Pa.}
\label{setup}
\end{figure}

\section{Experimental setup}
Figure~\ref{setup} illustrates the searching setup.
A linearly polarized creation beam (Ti:sapphire pulsed laser) and a circularly polarized 
inducing beam (Nd:YAG pulsed laser) were combined with a dichroic mirror (DM2) 
in the transport chamber by sharing a common optical axis.
At P1 in advance of the pulse compression,
the linear polarization state of the creation laser was introduced.
To transmit only P-polarized waves, P1 was made of 30 synthetic quartz plates tilted at Brewster's angle.
The measured extinction ratio with the full energy shots was approximately 1000 (P-pol.) : 1 (S-pol.).
The inducing laser was initially produced at P2 as a linearly polarized beam
with a commercial polarization beam splitter with an extinction ratio of 200 (P-pol) : 1 (S-pol.), 
and then the linear polarized state was further converted into the circular polarization state
by a quarter-plate ($\lambda /4$).
 
The central wavelengths of the creation and inducing lasers were 816 and 1064~nm, respectively,
their pulse durations were 31~fs and 9~ns, respectively, and 
their beam diameters were 37 and 
16~mm, respectively.
Inside the interaction chamber, the two incident beams were focused simultaneously in the vacuum 
by an off-axis parabolic mirror (OAP1) with a focal length of 279.1~mm. 
Based on an infinity-corrected optical system,
focal-spot images of the two beams were transferred to a CCD camera outside the interaction chamber.
The camera recorded the number of beam photons per pixel with a spatial resolution of 0.3~$\mu$m. 
A thin mesh with a known physical size located at the interaction point (IP) was used 
to calibrate the physical image size on the CCD camera.
By adjusting optical components inside the transport
chamber for the creation laser and ones outside the chamber for the inducing laser,
spatial overlap was ensured based on the focal spot images of the two beams.
The symmetrically placed identical off-axis parabolic mirror (OAP2) located at the point subtending the IP
collected the signal waves by restoring the plane-wave propagation of the two incident beams.
The intense incident beams were dumped through a dichroic mirror (DM3) 
that allowed the two incident beams to pass through while reflecting the signal waves.

With respect to the central wavelengths $\lambda_c$ and 
$\lambda_i$ for the creation and inducing lasers, respectively, 
the central wavelength $\lambda_s$ of the signal is defined by 
$\lambda_s=(2/\lambda_c-1/\lambda_i)^{-1} = 651$~nm.
This wavelength is close to the 633~nm of a He:Ne laser, so 
a He:Ne laser was combined at DM1 and DM2 with the inducing and creation lasers, respectively, 
as a calibration source, by which one can trace the signal trajectory down to the signal detector.
This calibration laser was used for aligning all the optical components inside the interaction chamber,
and it also had a role in evaluating the acceptance factor from the IP to the detector
with the specified circular polarization state.


The dichroic mirrors DM3--DM7 were custom-made and commonly reflected 651~nm with 99\% while transmitting around 816~nm with 99\% 
and 1064~nm with 95\%, to pick up the signal waves among the residual creation and inducing laser beams. 
To provide timing signals to synchronize the creation and inducing pulses
and also to monitor the stability of the pulse energies,
photodiodes (PD1, PD2) with a time resolution of $\sim$40~ps were used
by sampling the attenuated combined pulses that had passed through DM4.

A signal detector was used in the form of an R7400-01 single-photon-countable photomultiplier tube (PMT) 
manufactured by HAMAMATSU. The falling time resolution was 0.75~ns, which is
close to the waveform sampling resolution of 0.5~ns as explained below.
For the signal wavelengths of 610--690~nm to be selected,
a set of a low-pass filters transmitting above 610~nm and 
three types of band-pass filters transmitting 570--800~nm, 500--930~nm, and 450--690~nm
were installed in front of the PMT to remove residual photons from the intense incident lasers.

The voltages from the PMT and the two PDs as functions of time were recorded
using a waveform digitizer with a time resolution of 0.5~ns. 
The digitizer was triggered by a basic 10-Hz laser oscillator clock
to which the incident timing between the creation and inducing lasers was synchronized.
The incident rate of the creation laser was reduced to equi-interval 5~Hz by a mechanical shutter (MS1),
whereas that of the inducing laser was adjusted to non-equi-interval 5~Hz by a mechanical shutter (MS2) 
to produce four staggered trigger patterns for the offline waveform analysis.
The four types of trigger were as follows:
(i) two-beam incidence, ``S"; (ii) only inducing-laser incidence, ``I"; 
(iii) only creation-laser incidence, ``C"; (iv) no beam incidence, ``P".
These were issued in order over a data acquisition run, which ensured
equal shot statistics per trigger pattern and also minimized the systematic
uncertainties associated with subtractions between trigger patterns as explained
in the next section.

\section{Data analysis}
\subsection{Counting number of photons by means of a peak finder}
\begin{figure}[!h]
\begin{center}
\includegraphics[scale=0.8]{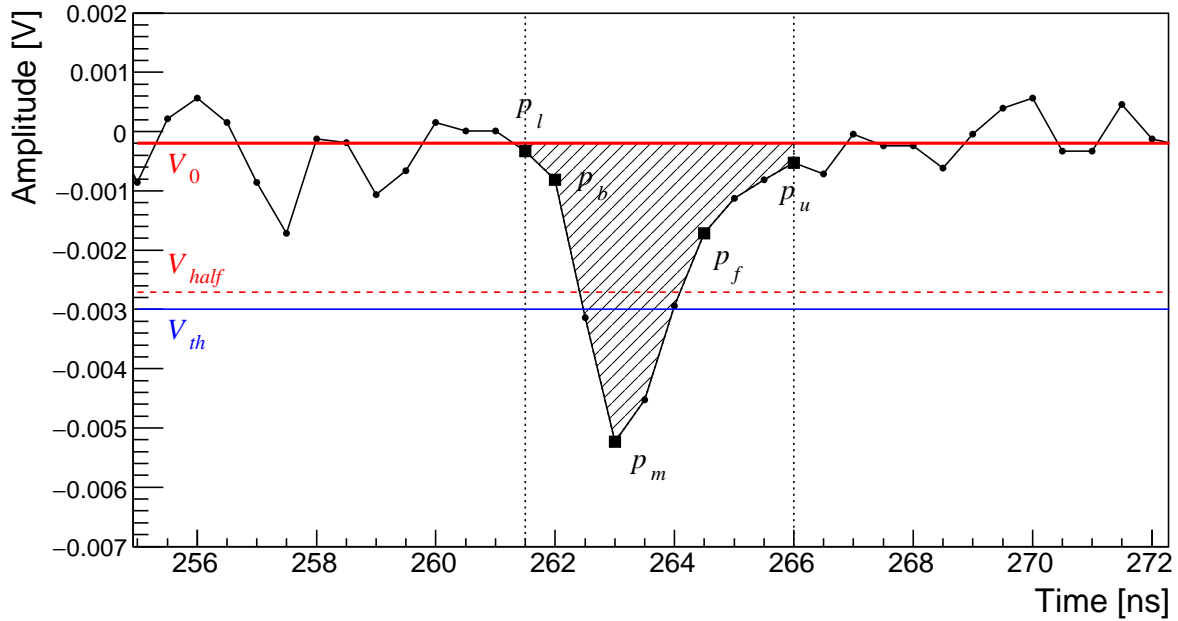}
\end{center}
\caption{Waveform sample including a peak with a single trigger. The shaded area shows the integral range used to 
evaluate the charge sum of the peak structure.}
\label{wfm_sample}
\end{figure}

The number of photons was evaluated based on the digitized waveform data from the PMT.
In a waveform (sometimes referred to as a shot), voltage values $V_i$ were recorded with respect to
individual sampling point $i$ within a 500-ns time interval as shown in Fig.~\ref{wfm_sample}. 
Because the time resolution (i.e., the width of a time bin) is 0.5~ns, $i$ runs from 1 to 1000.
A peak finder identifies peak structures in a waveform on a shot-by-shot basis and counts
the number of observed photons by the following steps:
(i) the finder determines an average voltage $V_0$ over $i=1$ to $400$;
(ii) by setting a proper threshold voltage $V_{th}$, the finder identifies a minimum voltage $V_m$ at $i=m$
in the peak-like domain above $V_{th}$;
(iii) a half-voltage value is defined as $V_{half}=(V_m + V_0)/2$;
(iv) along the forward (f) and backward (b) directions in time, the finder determines time points
$i=f$ and $i=b$ at which voltage values exceed $V_{half}$;
(v) from $i=f, b$, the falling $(i=l)$ and rising $(i=u)$ edges of a peak are defined as
$l = m - (2(m-b)-1)$ and  $u = m + 2(f-m)$, respectively;
(vi) a charge sum $Q$ contained in the peak-like structure depicted as the shaded area 
in Fig.~\ref{wfm_sample} is evaluated as
$Q = \sum_{i=l}^{u} V_i \Delta t/R$
with $\Delta t = 0.5$~ns and $R=50$~$\Omega$;
(vii) given a single-photon equivalent charge $Q_{p.e.} = (8.206 \pm 0.015) \times 10^{-14}$~C calibrated in advance, 
$Q/Q_{p.e.}$ provides the observed number of photons in the peak-like structure.

\begin{figure}[!h]
\begin{center}
\includegraphics[scale=0.8]{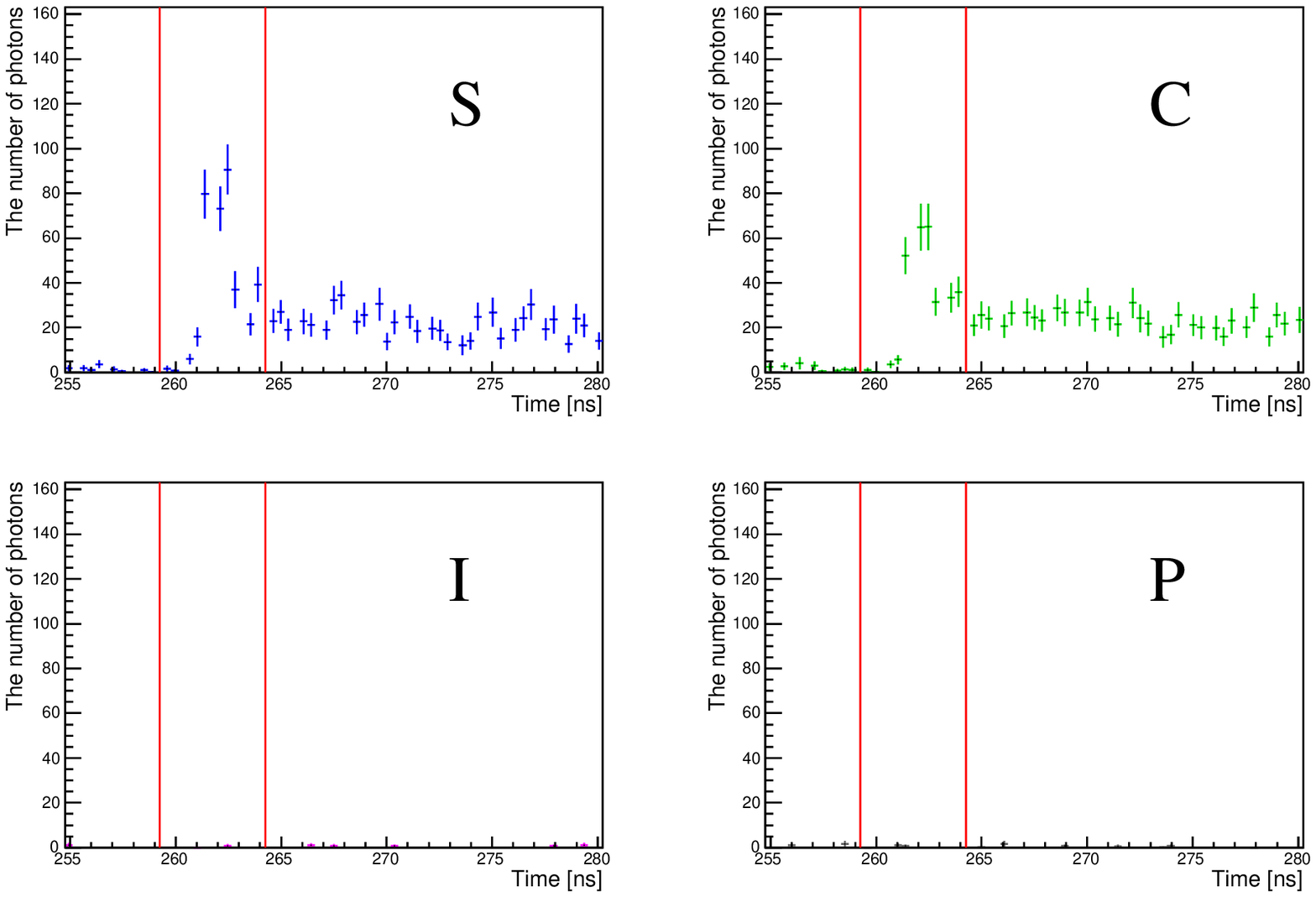}
\end{center}
\caption{Arrival-time distributions of number of observed photons by 
combining the P-polarization state (creation) and the left-handed circular polarization state (inducing)
for individual trigger patterns at 10~Pa. 
The individual time windows in which signal photons are expected to be detected 
are indicated with the two red lines.}
\label{peak_atm}
\end{figure}
 
\subsection{Pressure dependence of atomic four-wave mixing process}
Nonlinear optical parametric effects caused by third-order polarization susceptibility $\chi^{(3)}$,
so-called four-wave mixing (FWM)~\cite{FWM}, are expected to occur even in the residual gas in the interaction 
vacuum chamber. This can be a dominant background source because the wavelengths generated in atomic FWM are 
nearly equal to those of stimulated scattering in a vacuum due to the kinematic similarity 
based on energy--momentum conservation between the initial and final state photons.
Therefore, we can refer to the searched-for stimulated scattering process as FWM in a vacuum.
Meanwhile, atomic FWM is quite useful for ensuring spatiotemporal synchronization
between the creation and inducing pulse lasers in QPS.
To validate our searching system, we measured the pressure dependence of the number of atomic FWM photons.
Because the peak finder can provide the falling edges of photon incident peaks in the waveforms, 
the arrival-time distribution as shown in Fig.~\ref{peak_atm} is measurable in units of the number of photons
for each of the four trigger patterns. This figure shows the results measured at 10~Pa.
The peak structures appeared in trigger patterns~S and C. The peak seen in pattern~C is 
expected because of plasma creation at the IP because the creation laser intensity is high enough 
to induce ionization of residual atoms. In contrast, the intensity of the inducing laser field 
is much lower because of the long time duration, as seen in pattern~I where no peak is found.
Meanwhile, the higher peak seen in pattern~S is expected to be the sum between 
the atomic FWM and the plasma-origin photon yields.
\begin{figure}[!h]
\begin{center}
\includegraphics[scale=0.7]{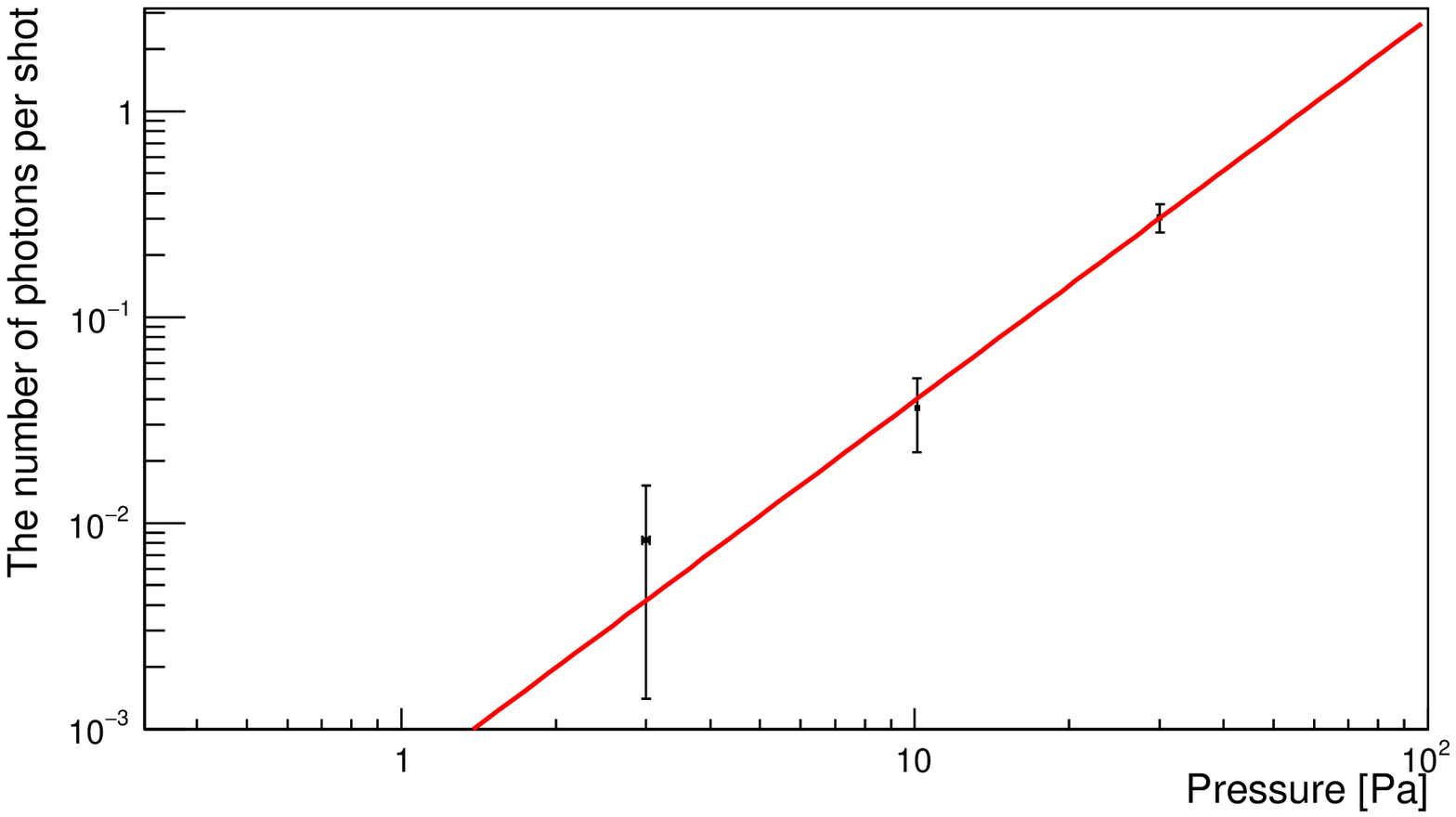}
\end{center}
\caption{Pressure dependence of number of four-wave-mixing photons per shot
from residual atoms inside interaction chamber
when P-pol.\ (creation) and left-handed circular (inducing) polarization laser pulses 
are combined and focused.}
\label{Fig7}
\end{figure}

The basic assumption that addition of the number of photons in individual trigger patterns 
corresponds to the number of photons in trigger pattern~S is indeed supported by the following subtraction analysis.
The acceptance-uncorrected number of atomic FWM photons, $N_{S}$, can be obtained via
\beq
N_{S} = (n_{S}-n_{P}) - (n_{C}-n_{P}) - (n_{I}-n_{P}) = n_{S} - n_{C} - n_{I} + n_{P},
\label{sub}
\eeq
where $n_{i}$ is the number of photons for trigger pattern $i$
measured in the time interval subtended by the two red vertical lines 
corresponding to the signal generation timing window.
Figure~\ref{Fig7} shows the pressure dependence of the number of signal photons per shot,
which is expected to be dependent upon the square of the pressure because the photon yield of the atomic FWM
should be proportional to $(\chi^{(3)})^2 \propto (density)^2 \propto (pressure)^2$.
The dependence was thus fitted with 
\beq\label{eq_scaling}
N_{S}/shot = a {\cal P}^{b},
\eeq
where $a$ and $b$ are fitting parameters and ${\cal P}$ is pressure.
The error bars are the quadratic sum of the statistical error propagation associated 
with the subtraction process between trigger patterns
and systematic uncertainties of focal-point stability during a run period.
We explain these uncertainties in the following subsection.
As expected, $b = 1.85 \pm 0.35$ is close to the expected behavior $N_S \propto {\cal P}^2$ 
in atomic physics~\cite{FWM, PTEP-EXP01}.
Note that this pressure dependence itself is valuable as data because
the special combination between linear and circular polarization state beams 
is a very rare case in atomic physics.

\subsection{Focal-point stability}
\begin{figure}[!h]
\begin{center}
\includegraphics[scale=1.00]{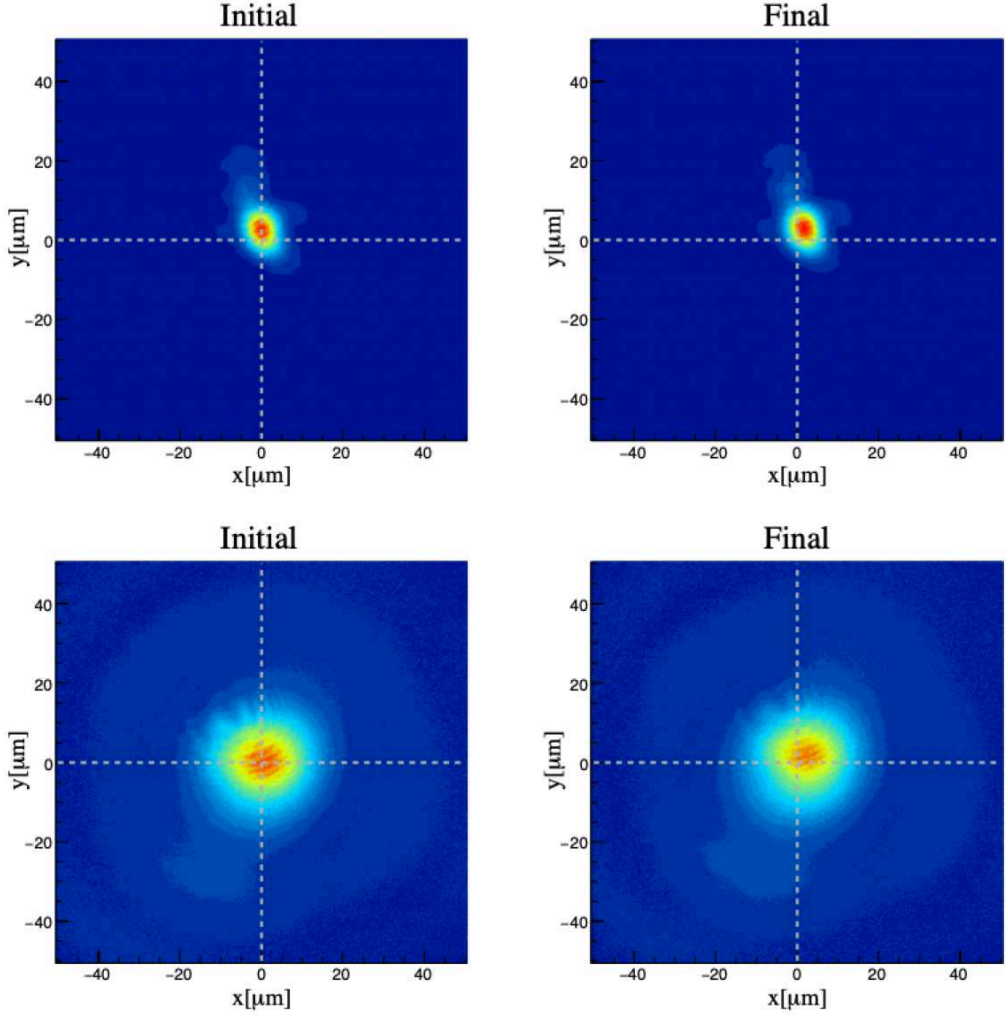}
\end{center}
\caption{Beam profiles of creation (upper) and inducing (lower) lasers at interaction point as captured
by a common CCD camera. The left and right figures correspond to typical images taken at
the beginning and end of a unit run period, where slight deviations in the focal spots are seen.}
\label{profile}
\end{figure}

The systematic uncertainties due to focal-point fluctuations
were estimated from overlaps between the two laser focal-spot profiles measured 
by the common CCD camera sensitive to both wavelengths.
Figure~\ref{profile} shows typical focal-spot images of the creation and inducing lasers.
%

With the local intensity per CCD pixel of the monitor camera, $N(x,y)$,
the overlap factor $O$ is defined as
\beq\label{eq_O0}
O \equiv \sum_{x}^{c} \sum_{y}^{c} N^2_c(x,y) N_i(x,y),
\eeq
where the subscripts $c$ and $i$ specify the creation and inducing lasers, respectively.
The summations were taken over the area framed by the full width at half maximum of the creation laser intensity profile.
Fluctuations of the overlap factors with respect to the mean overlap factor $(O_I+O_F)/2$
were then evaluated as 
\beqa
\delta N_{S} = \left|N_{S} \frac{O_I - O_F}{O_I + O_F}\right|,
\label{eq_O}
\eeqa
where $O_{I,F}$ are the overlap factors at the beginning and end, respectively,
of a 2000-s unit run period.

\begin{figure}[!h]
\begin{center}
\includegraphics[scale=0.8]{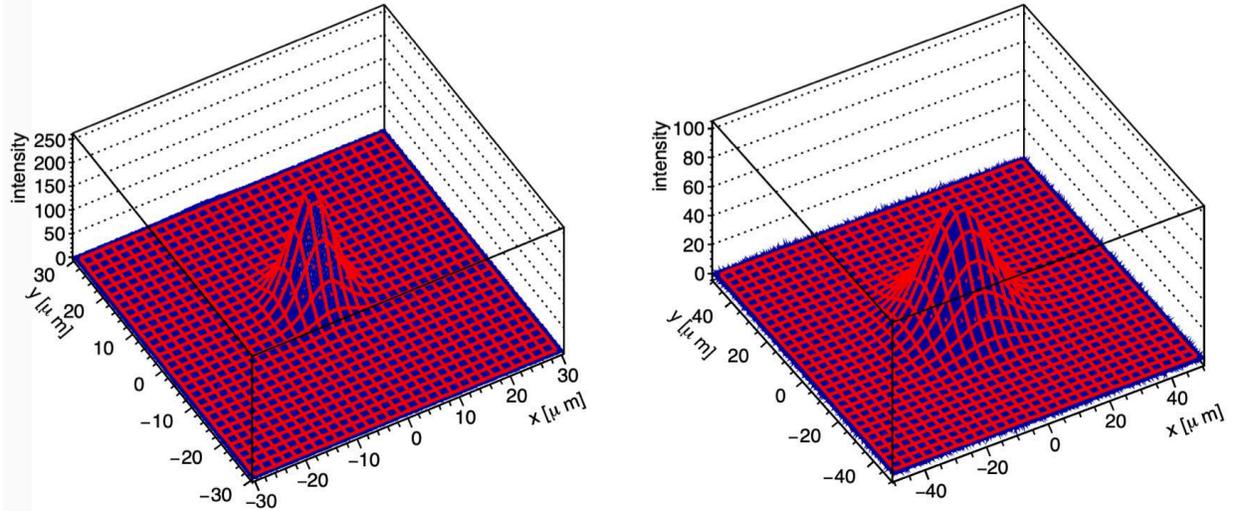}
\end{center}
\caption{
Fitting of focal-plane intensity profiles with two-dimensional Gaussian distributions constrained by $x$--$y$ symmetry
for creation (left) and inducing (right) beams.
}
\label{ProfileFit}
\end{figure}

\subsection{Effective energy fraction in Gaussian beams}
Figure~\ref{ProfileFit} shows the results of fitting the focal-plane intensity profiles of the creation and inducing beams with two-dimensional Gaussian distributions constrained by $x$--$y$ symmetry.
From the fitting results $\sigma_{xy} = 7$ and $17$~$\mu$m for the
creation and inducing lasers, respectively, we evaluated the effective energy fraction
contained in the region within 3~$\sigma_{xy}$
among the entire intensity profiles, including the peripheral diffraction parts
that are assumed not to contribute to stimulated photon--photon scattering.

\section{Search result}
\begin{figure}[!h]
\begin{center}
\includegraphics[scale=0.8]{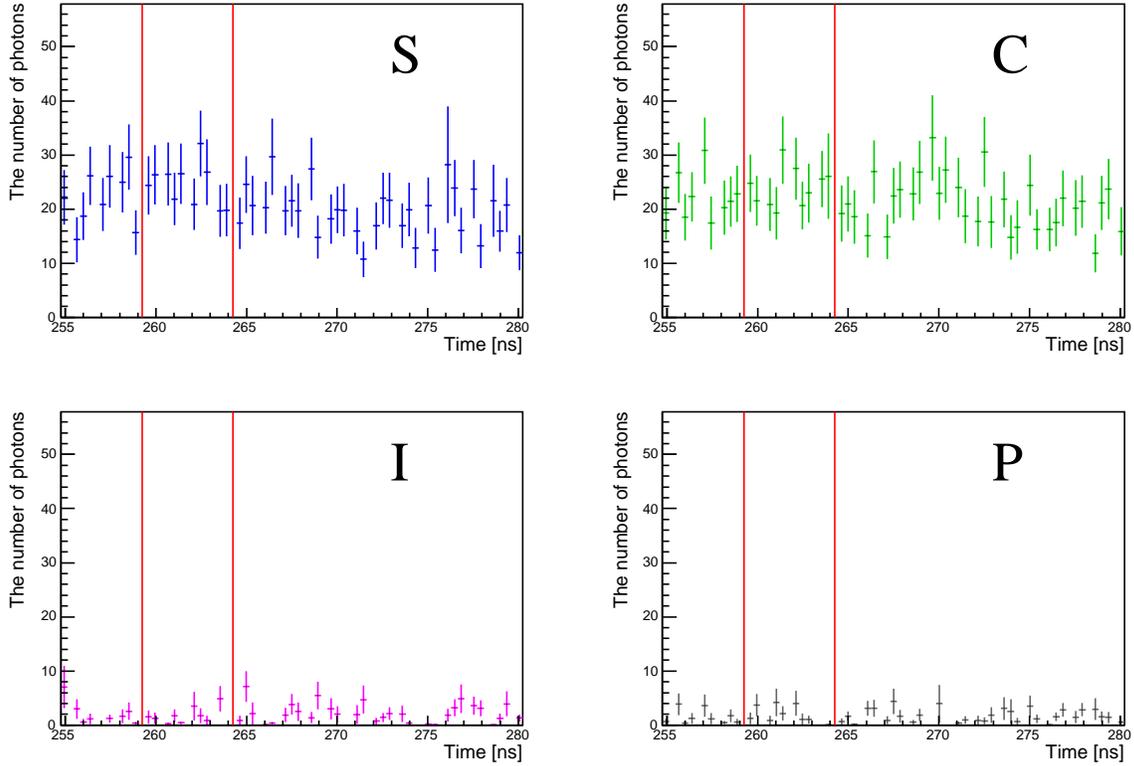}
\end{center}
\caption{Arrival-time distributions of detected photons at $2.6\times10^{-5}$~Pa
for trigger patterns~S, C, I, and P by combining the P-polarization state (creation) and
the left-handed circular polarization state (inducing).}
\label{Fig11}
\end{figure}

By combining linearly polarized creation laser pulses and circularly polarized inducing laser pulses,
searches for scalar and pseudoscalar resonance states were performed
at a vacuum pressure of $2.6 \times 10^{-5}$~Pa.
The arrival-time distributions of photons identified by the peak finder are shown in Fig.~\ref{Fig11}
for the individual trigger patterns. 
The area subtended by the two red lines corresponds to the expected signal timing window. 
The total number of shots in trigger pattern~S was $W_S \equiv 2.9993 \times 10^4$.

\begin{figure}[!h]
\begin{center}
\includegraphics[scale=0.70]{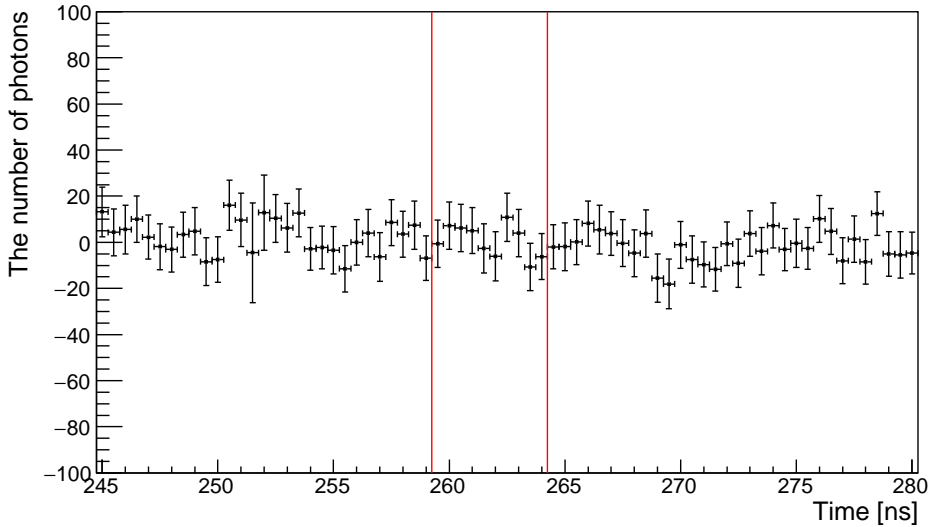}
\end{center}
\caption{Arrival-time distribution of number of photons obtained by applying Eq.~(\ref{sub})
to the entire timing windows including the signal window enclosed by the two red lines
at $2.6\times10^{-5}$~Pa by combining the P-polarization state (creation) and
the left-handed circular polarization state (inducing).}
\label{Fig12}
\end{figure}

Figure~\ref{Fig12} shows the arrival-time distribution of the number of photons after 
subtraction between different trigger patterns based on Eq.~(\ref{sub}).
The total number of signal photons within the signal incident timing window was obtained as
\beq
N_{S} = 4.9 \pm 22.8 (\rm{stat.}) \pm 22.8 (\rm{syst.I}) \pm 3.8 (\rm{syst.I\hspace{-1pt}I}) \pm 3.7 (\rm{syst.I\hspace{-1pt}I\hspace{-1pt}I}).
\label{result}
\eeq
Systematic error~I was estimated by measuring the root-mean-square of 
the number of photon-like signals excluding the signal window,
corresponding to the baseline uncertainty on the number of photons.
Systematic error~II reflects time variations of the overlap factors defined in Eq.~(\ref{eq_O}) between
the focal spots of the creation and inducing lasers.
Note that this equation contains fluctuations of beam energies during a run period
as well as the pointing fluctuations.
%
Systematic error~III was obtained by changing the threshold value in the peak finder, $V_{th} = (-1.3 \pm 0.1)$~mV,
by assuming a uniform distribution from $-1.2$ to $-1.4$~mV.

\section{Upper limits on coupling--mass relation for ALP exchanges}
From the result in (\ref{result}), we conclude that
no signal photons in the quasi-vacuum state were observed based on the total uncertainty.
Indeed, this result is also consistent with the expected number of background photons 
per shot (efficiency-uncorrected) due to residual gases, estimated as
\beq\label{eq_Ngas}
N_{gas} / shot = 1.8 \times 10^{-12} \quad \mbox{photons}
\eeq
by extrapolating to $2.6 \times 10^{-5}$~Pa with Eq.~(\ref{eq_scaling}).
In addition, for the given total statistics, 
the expectation value based on the QED photon--photon scattering process, 
which is the only possible process in the standard model, is negligibly low
at $E_{cms} < 1$~eV~\cite{PTEPGG} even though the stimulation effect is taken into account~\cite{PTEP17}.
Therefore, with respect to a null hypothesis following a Gaussian distribution,
we provide the upper limits on the coupling--mass relation by
assuming scalar and pseudoscalar field exchanges with the experimental parameters in Table~\ref{Tab1}.

We note that the pulse duration of the Nd:YAG laser, $\tau_{ibeam}$, in Table~\ref{Tab1} 
is not corresponding to that of the Fourier transform limit
due to the different scheme to generate pluses from that of Ti:sapphire laser 
in which time duration close to reaching the Fourier transform limit is obtained.
Thus, the effective time duration reaching the Fourier transform limit, $\tau_i$, which can overlap 
with the creation pulse duration, $\tau_c$, is evaluated from the spectrum linewidth of the Nd:YAG laser. 
This treatment is consistent with the basic assumption in~\cite{JHEP} where the inducing effect is evaluated
based on overlapping pulses individually reaching Fourier transform limits.
In addition to the effective time durations, by considering the spatially overlapping regions 
within 3~$\sigma_{xy}$ focal spots which are consistent with the Gaussian shapes, 
the effective numbers of photons per pulses, $N_c$ and $N_i$, were used for 
the following limit calculations. 
In this sense, the following results correspond to conservative upper limits,
because the effective beam energies stored in pulses are very much reduced.

As for the upper mass range, because of the inclusion of general asymmetric collisions, 
this search is sensitive to a heavier mass range compared with the symmetric collision range, expressed as
\beq
m = 2\omega_c\sin\Delta\theta \sim 2\omega_c \frac{d_c}{2f} = 0.21 \mbox{~eV}
\eeq
based on values in Table~\ref{Tab1} with $\Delta\theta \equiv d/(2f)$ defined 
by the focal length $f$ and beam diameter $d$ of the creation laser in Fig.~\ref{Fig1}.
Note, however, that this value is merely a reference mass at which the maximum sensitivity is expected.

\begin{figure}[!h]
\begin{center}
\includegraphics[scale=0.80]{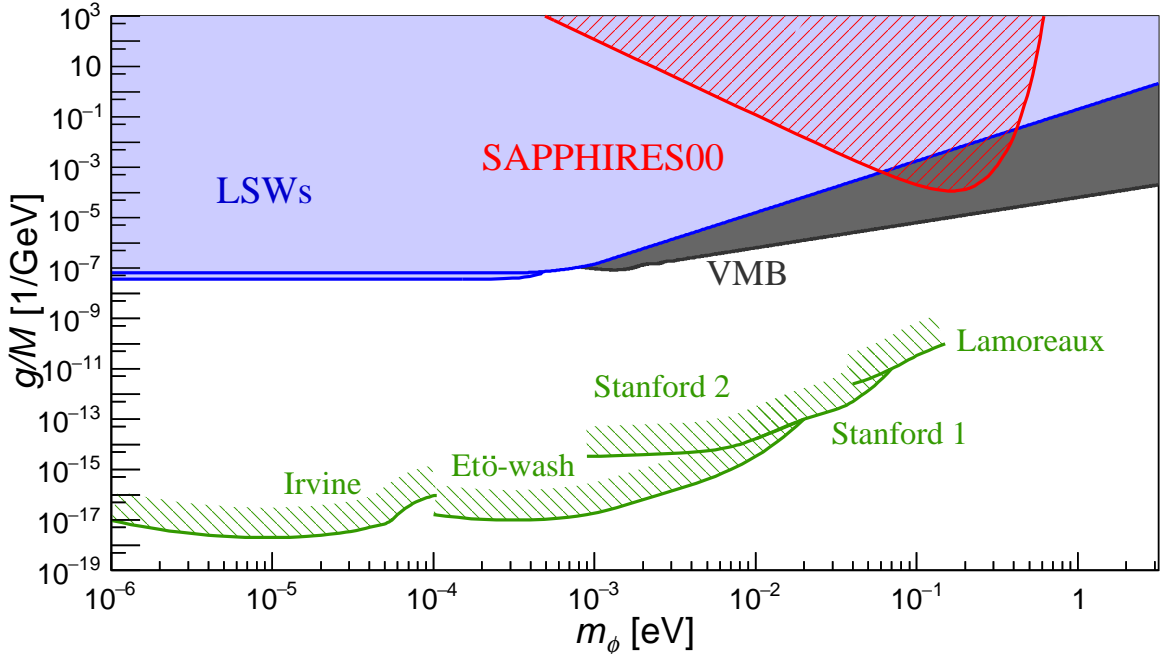}
\end{center}
\caption{Upper limits on coupling--mass relation for scalar field ($\phi$) exchanges.
The red shaded area labeled "SAPPHIRES00" is the region excluded by this work (stimulated resonant scattering). The light-blue area is the region for scalar fields excluded
by the "Light Shining through a Wall experiments (LSWs)" (OSQAR~\cite{osqar} and ALPS~\cite{alps})
with simplification of the sine-function part to unity above $10^{-3}$~eV for drawing convenience.
The gray area is the result from the "Vacuum Magnetic Birefringence (VMB)" experiment (PVLAS~\cite{pv}). 
The green shaded areas are regions excluded based on non-Newtonian force searches 
("Irvine"~\cite{Irvine}, "Eto-wash"~\cite{Eto-wash}, "Stanford1"~\cite{Stanford1}, "Stanford2"~\cite{Stanford2}) and on Casimir force measurements ("Lamoreaux"~\cite{Lamoreaux}).
}
\label{coupling_sc}
\end{figure}

\begin{figure}[!h]
\begin{center}
\includegraphics[scale=0.80]{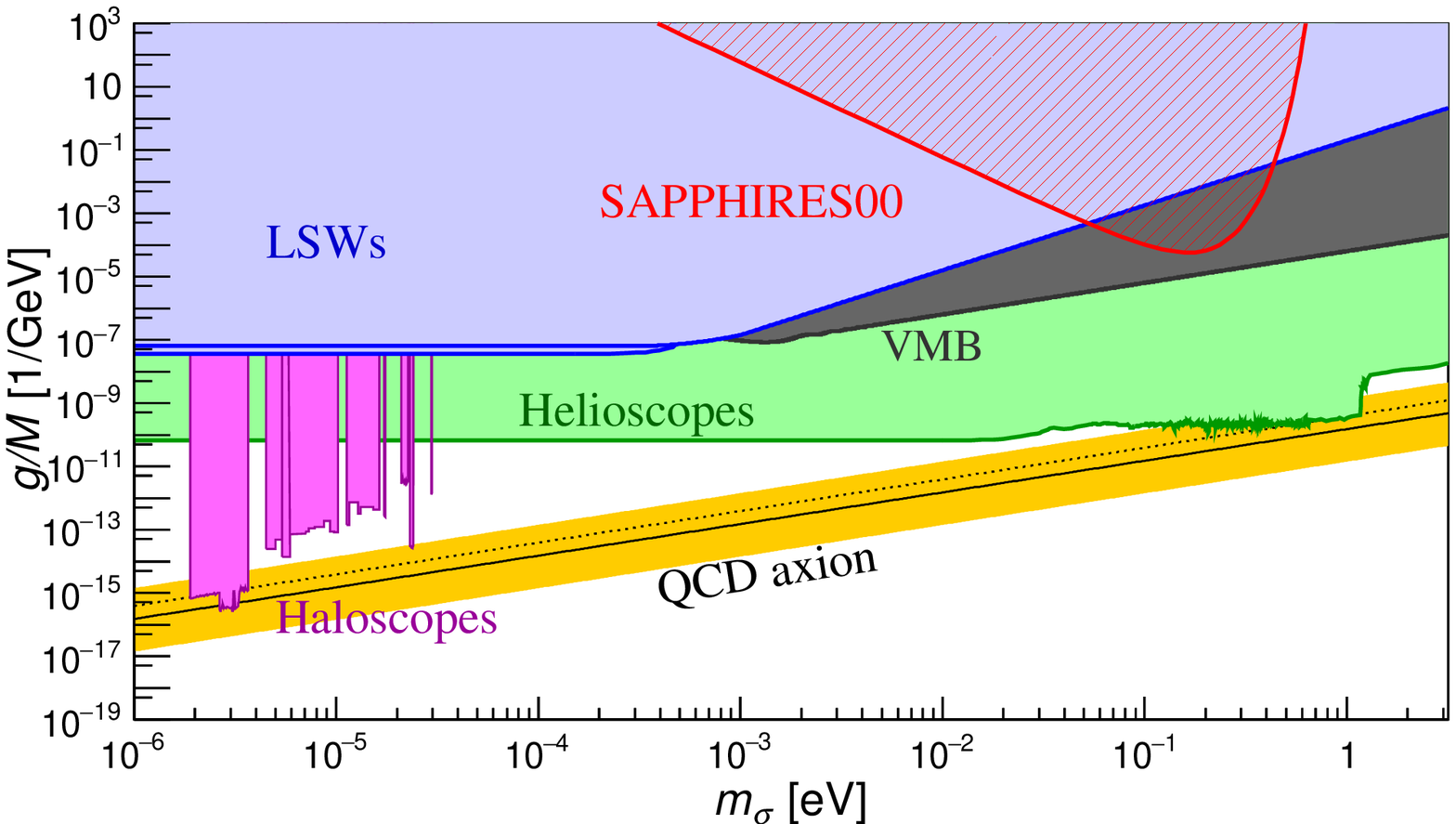}
\end{center}
\caption{
Upper limits on coupling--mass relation for pseudoscalar fields ($\sigma$) exchanges.
The red shaded area labeled "SAPPHIRES00" is the region excluded by this work
(stimulated resonant scattering).
The yellow band shows the coupling--mass relation based on the QCD axion predicted 
by the KSVZ model~\cite{KSVZ} with $|E/N-1.95|$ in the range 0.07--7
with the case of $E/N=0$ (black dotted line).
The case of the DFSZ model~\cite{DFSZ} with $E/N=8/3$ is also shown by the black solid line.
The light-blue area is the region excluded by LSW experiments (OSQAR~\cite{osqar} and ALPS~\cite{alps}) 
with respect to pseudoscalar fields.
The gray area is the result from the "Vacuum Magnetic Birefringence (VMB)" experiment (PVLAS~\cite{pv}). 
The green area shows upper limits from the "Helioscope" experiment (CAST~\cite{cast}).  
The magenta areas are excluded regions from the "Haloscope" experiments (ADMX~\cite{admx}, RBF~\cite{rbf}, UF~\cite{uf}, and HAYSTAC~\cite{haystac}). 
}
\label{coupling_ps}
\end{figure}

A confidence level $1-\alpha$ to exclude a null hypothesis is expressed as 
\beq
1-\alpha = \frac{1}{\sqrt{2\pi}\sigma}\int^{\mu+\delta}_{\mu-\delta}
e^{-(x-\mu)^2/(2\sigma^2)} dx = \mbox{erf}\left(\frac{\delta}{\sqrt 2 \sigma}\right),
\eeq
where $\mu$ is the expected value of an estimator $x$ following a hypothesis, and 
$\sigma$ is one standard deviation.
In this search, the estimator $x$ corresponds to $N_{S}$, and
we assign the acceptance-uncorrected uncertainty $\delta N_{S}$ 
from the quadratic sum of all error components in the result (\ref{result})
as the one standard deviation $\sigma$ around the mean value $\mu=0$.
In this search, the null hypothesis is produced from fluctuations of the number of photon-like signals
following a Gaussian distribution whose expectation value, $\mu$, is zero 
for the given total number of shots, $W_S=2.9993 \times 10^4$.
This is because $N_S$ is calculated from subtractions between different trigger patterns
whose baseline fluctuations, in principle, should follow Gaussian distributions individually.
To obtain a confidence level of 95\%,
$2 \alpha = 0.05$ with $\delta = 2.24 \sigma$ is used, where
a one-sided upper limit by excluding above $x+\delta$~\cite{PDGstatistics} is applied.
To evaluate the upper limits on the coupling--mass relation,
we then solved
\beq
2.24 \delta N_S = {\cal Y}_{c+i}(m, g/M ; P) t_{a} r \epsilon
\eeq
numerically based on Eq.~(\ref{Nobs}) with respect to $m$ and $g/M$ for a set of 
experimental parameters $P$ in Table~\ref{Tab1}, where $t_a r = W_S = 2.9993 \times 10^4$ and
the overall efficiency $\epsilon \equiv \epsilon_{opt}\epsilon_d $
with the optical path acceptance $\epsilon_{opt}$ to the $p_3$ detector position
and the single $p_3$-photon detection efficiency $\epsilon_d$ were used.
Figures~\ref{coupling_sc} and \ref{coupling_ps} show the obtained upper limits
on the coupling--mass relations
for scalar and pseudoscalar fields, respectively, at a 95\% confidence level.
Note that based on Eq.~(\ref{helicity}), we used
$\epsilon_{opt}=\epsilon_{L}$ as the optical path acceptance factor, where $\epsilon_{L}$ is
the acceptance factor with respect to left-handed circularly polarized photons
measured from the IP to the $p_3$-detection position.
This is because both scalar and pseudoscalar fields
can couple only to the same helicity state as that of the inducing field, 
which is provided as the left-handed state in the searching setup.

\begin{table}[h!]
\caption{Experimental parameters used to numerically calculate the upper limits on the coupling--mass relations.
The effective numbers of photons, $N_c$ and $N_i$, were used for the limit calculations.
}
\begin{center}
\begin{tabular}{lr}  \\ \hline 
Parameter & Value \\ \hline
Centeral wavelength of creation laser $\lambda_c$   & 816 nm\\
Relative linewidth of creation laser, $\delta\omega_c/<\omega_c>$ &  $1.2\times 10^{-2}$\\
Duration time of creation laser, $\tau_{c}$ & 31 fs $( = 36\,\mbox{fs}(\mbox{FWHM})/\sqrt{2 \ln 2})$\\
Measured creation laser energy per $\tau_{c}$, $E_{c}$ & (100 $\pm$ 5) $\mu$J \\
Creation energy fraction within 3~$\sigma_{xy}$ focal spot, $f_c$ & 0.85\\
Effective creation energy per $\tau_c$ within 3~$\sigma_{xy}$ focal spot & $E_{c} f_c = 85$~$\mu$J\\
Effective number of creation photons, $N_c$ & $3.5 \times 10^{14}$ photons\\
Beam diameter of creation laser beam, $d_{c}$ & (37.0 $\pm$ 0.8)~mm\\
Polarization & linear (P-polarized state) \\ \hline
Central wavelength of inducing laser, $\lambda_i$   & 1064~nm\\
Relative linewidth of inducing laser, $\delta\omega_{i}/<\omega_{i}>$ &  $1.0\times 10^{-4}$\\
Duration time of inducing laser beam, $\tau_{ibeam}$ & 9~ns\\
Measured inducing laser energy per $\tau_{ibeam}$, $E_{i}$ & $(200 \pm 4)$~$\mu$J \\
Linewidth-based duration time of inducing laser, $\tau_i/2$ & $\hbar/(2\delta\om_{i})=2.8$~ps\\
Inducing energy fraction within 3~$\sigma_{xy}$ focal spot, $f_i$ & 0.87\\
Effective inducing energy per $\tau_i$ within 3~$\sigma_{xy}$ focal spot & $E_{i} (\tau_i/\tau_{ibeam}) f_i = 0.11$~$\mu$J\\
Effective number of inducing photons, $N_i$ & $5.9 \times 10^{11}$ photons\\
Beam diameter of inducing laser beam, $d_{i}$ & $(15.8 \pm 0.3)$~mm\\
Polarization & circular (left-handed state) \\ \hline
Focal length of off-axis parabolic mirror, $f$ & 279.1~mm\\
Single-photon detection efficiency, $\epsilon_{d}$ & 1.4\% \\
Efficiency of optical path from IP to PMT, $\epsilon_{L}$ & 33\% \\ \hline
Total number of shots in trigger pattern S, $W_S$   & $2.9993 \times 10^4$ shots\\
$\delta{N}_{S}$ & 32.7\\
\hline
\end{tabular}
\end{center}
\label{Tab1}
\end{table}

\section {Conclusions} 
By combining linearly polarized creation laser pulses and circularly polarized inducing laser pulses,
we have searched for scalar and pseudoscalar fields via stimulated resonant scattering
by focusing two-color pulsed lasers: 0.10~mJ/31~fs at 816~nm and 0.20~mJ/9~ns at 1064~nm.  
The observed number of signal photons in the quasi-vacuum state at $2.6 \times 10^{-5}$~Pa was 
$4.9 \pm 22.8 (\rm{stat.}) \pm 22.8 (\rm{syst.I}) \pm 3.8 (\rm{syst.I\hspace{-1pt}I}) \pm 3.7 (\rm{syst.I\hspace{-1pt}I\hspace{-1pt}I})$.
We thus conclude that no significant signal was observed in this search. 
The expected number of signal photons from the residual gas is sufficiently low 
based on the upper limit from the measurement of the pressure dependence.  
Based on the assumption that uncertainties are dominated by systematic fluctuations
around the zero expectation value following a Gaussian distribution and 
the fully asymmetric collisional geometry in quasi-parallel stimulated photon--photon scattering,
we provided upper limits on the coupling--mass relations for scalar and pseudoscalar fields 
at a 95\% confidence level in the sub-eV mass range.

\section*{Acknowledgments}
K.\ Homma acknowledges the support of the Collaborative Research
Program of the Institute for Chemical Research of Kyoto University 
(Grant Nos.\ 2018--83, 2019--72, 2020--85, and 2021--88)
and Grants-in-Aid for Scientific Research
Nos.\ 17H02897, 18H04354, 19K21880, and 21H04474 from the Ministry of Education, 
Culture, Sports, Science and Technology (MEXT) of Japan.
The authors in ELI-NP acknowledge the support by Extreme Light Infrastructure Nuclear Physics Phase II, 
a project co-financed by the Romanian Government and the European Union 
through the European Regional Development Fund and 
the Competitiveness Operational Programme (No. 1/07.07.2016, COP, ID 1334). 

\newpage
\section*{Appendix}
In order to configure for the actual experimental condition where two beam diameters are different,
we replace $\mcal{D}_I$ in Eq.(A.84) of \cite{JHEP} prepared for the application to the common diameter case
between creation (subscript $c$) and inducing (subscript $i$) beams with the following factor $\mcal{D}_{c+i}$
by taking the diameter difference into account. 
\begin{equation} \label{Dci}
\mcal{D}_{\mrm{c+i}}
= \frac{1}{2}\left(\frac{2}{\pi}\right)^{\frac{3}{2}}\frac{1}{c^{2}}
  \frac{\tau_{\mrm{i}}}{\tau_{\mrm{c}}}\frac{1}{\sqrt{\tau_{\mrm{c}}^{2}+2\tau_{\mrm{i}}^{2}}}
  \frac{1}{w_{\mrm{c}0}^{2}\left(1-\frac{z_{\mrm{c}R}^{2}}{z_{\mrm{i}R}^{2}}\right)}
  \left[ z_{\mrm{c}R}\tan^{-1}\left( \frac{z_{\mrm{i}R}}{z_{\mrm{c}R}} \right)
  - R_{\mrm{ci}} Z_{\mrm{ci}}\tan^{-1}\left( \frac{z_{\mrm{i}R}}{Z_{\mrm{ci}}} \right) \right]
\end{equation}
with
\begin{equation}
R_{\mrm{ci}} \equiv \frac{w_{\mrm{c}0}^{2} z_{\mrm{i}R}^{2} + 2w_{\mrm{i}0}^{2} 
z_{\mrm{c}R}^{2}}{z_{\mrm{i}R}^{2}\left(w_{\mrm{c}0}^{2}+2w_{\mrm{i}0}^{2}\right)}
\end{equation}
and
\begin{equation}
\quad Z_{\mrm{ci}} \equiv \sqrt{\frac{w_{\mrm{c}0}^{2}+2w_{\mrm{i}0}^{2}}{\vth_{\mrm{c}0}^{2} + 2\vth_{\mrm{i}0}^{2}}}
\end{equation}
using beam diameters $d_{\mrm{k}}$, beam waists $w_{\mrm{k}0}$, and Rayleigh lengths $z_{\mrm{k}R}$ for $\mrm{k = c,\,i}$
defined as
\begin{align}
\vth_{\mrm{k}0} & = \tan^{-1}\left(\frac{d_{\mrm{k}}}{2f}\right), \\ 
w_{\mrm{k}0} & = \frac{\lambda_{\mrm{k}}}{\pi \vth_{\mrm{k}0}}, \\
z_{\mrm{k}R} & = \frac{\pi w_{\mrm{k}0}^{2}}{\lambda_{\mrm{k}}}.
\end{align}
We note that $\mcal{D}_{\mrm{c+i}}$ is obtained by integrating the spatiotemporal overlapping factor in Eq.(\ref{eq_Yci})
over the Rayleigh length of the inducing laser which is longer than that of the creation laser
in the experimental setup.

\end{document}